\documentclass{aa}  

\usepackage[varg]{txfonts}

\usepackage{graphicx}	
\usepackage{amsmath}	
\usepackage{amssymb}	
\usepackage{multicol}	
\usepackage{bm}		
\usepackage{pdflscape}	
\usepackage{ulem} 
\usepackage{lineno} 
\usepackage{multirow}

\usepackage[usenames,dvipsnames]{color}
\usepackage[hyperfootnotes=false]{hyperref}
\hypersetup{
  colorlinks,
  citecolor=Blue,
  linkcolor=Blue,
  urlcolor=Blue}  
  
\usepackage{txfonts}
\begin{document} 

  \titlerunning{Unified origin of galaxy scaling laws}
   \title{A unified scenario
  for the origin of spiral and elliptical galaxy structural scaling laws}

   
   \author{\href{https://orcid.org/0000-0002-1295-1132}{Ismael Ferrero}\inst{1,}\thanks{s.i.ferrero@astro.uio.no} \and \href{https://orcid.org/0000-0003-3862-5076}{Julio F. Navarro} \inst{2} \and \href{https://orcid.org/0000-0003-3055-6678}{Mario G. Abadi}\inst{3,4} \and \href{https://orcid.org/0000-0003-1896-0424}{Jos\'e A. Benavides}\inst{3} \and \href{https://orcid.org/0000-0003-0469-3193}{Dami\'an Mast}\inst{4,5}}

   \institute{Institute of Theoretical Astrophysics, University of Oslo.
P.O. Box 1029 Blindern, NO-0315 Oslo, Norway.
         \and
             Department of Physics \& Astronomy, University of Victoria,
 Victoria, BC, V8P 5C2, Canada.
          \and
             Instituto de Astronom\'ia Te\'orica y Experimental, CONICET-UNC, Laprida 854,
  X5000BGR, C\'ordoba, Argentina. 
          \and
             Observatorio Astron\'omico de Córdoba, Universidad Nacional de C\'ordoba,
 Laprida 854, X5000BGR, C\'ordoba, Argentina.  
            \and
             Consejo de Investigaciones Cient\'ificas y T\'ecnicas de la Rep\'ublica Argentina, Avda. Rivadavia 1917, C1033AAJ, CABA, Argentina.\\   
             }


  \abstract
   { Elliptical (E) and spiral (S) galaxies follow tight,
  but different, scaling laws that link their stellar masses, radii, and
  characteristic velocities. Mass and velocity, for example, scale
  tightly in spirals with little dependence on galaxy radius (the
  `Tully-Fisher relation'; TFR). On the other hand, ellipticals appear to trace
  a 2D surface in size-mass-velocity space (the `Fundamental Plane';
  FP).  Over the years, a number of studies have attempted to
  understand these empirical relations, usually in terms of variations
  of the virial theorem for E galaxies and in terms of the scaling
  relations of dark matter halos for spirals. We use Lambda cold dark matter ($\Lambda$CDM) cosmological hydrodynamical simulations to show that
  the scaling relations of both ellipticals and spirals arise as the
  result of (i) a tight galaxy mass--dark halo mass relation and (ii)
  the self-similar mass profile of cold dark matter halos. In this
  interpretation, E and S galaxies of a given stellar mass inhabit halos
  of similar masses, and their different scaling laws result from the
  varying amounts of dark matter enclosed within their luminous
  radii. This scenario suggests a new galaxy distance indicator
  applicable to galaxies of all morphologies and provides simple and
  intuitive explanations for long-standing puzzles, such as why the
  TFR is independent of surface brightness, or what causes the
  `tilt' in the FP. Our results provide strong support for the
  predictions of $\Lambda$CDM in the strongly non-linear regime,
  as well as guidance for further improvements to cosmological
  simulations of galaxy formation.}
   

\maketitle


\section{Introduction}

Understanding the origin of the scaling laws that relate the structural
parameters of galaxies has long been a key goal of galaxy formation
models. Success, however, has so far been elusive. This is due in part
to the complexity of the problem: Although the luminosity, the size, and
the characteristic velocity of galaxies are all strongly correlated, the
detailed relations depend on wavelength and differ in nature for
galaxies of different morphologies.

The wavelength dependence, however, has been sidestepped by
progress in our understanding of stellar evolution, which has made it
possible to combine luminosities in different bands to derive reliable
stellar mass ($M_*$) estimates for galaxies of different morphologies
\citep[see e.g.][]{Bell2001,Blanton2007}. In addition, the galaxy projected 
stellar half-mass radius, or `effective radius', $R_{\rm e}$, enables meaningful
comparison between the characteristic mass and size of galaxies of
widely different morphologies \citep[e.g.][]{Shen2003}.

Comparing the characteristic velocities of galaxies of different
morphologies is less straightforward. For late-type galaxies
(hereafter `spiral'; or `S', for short), where the luminous
(gas and stars) component is in a prominent disc, rotation velocities ($V_{\rm rot}$) are
amenable to observation, whereas in early-type galaxies
(hereafter `elliptical'; or `E'), which typically lack cold gas and a well-defined
disc component, the line-of-sight stellar velocity dispersion
($\sigma$) is the commonly adopted measure.

An added complexity is that these velocity measures typically depend
on the radius, although it is possible in general to identify a single 
characteristic value for each galaxy. Indeed,
rotation velocities in spirals vary little outside the very inner
regions \citep[i.e. rotation curves are `flat'; see e.g.][and
references therein]{Courteau2007}, and many ellipticals are well
approximated by `isothermal' models where $\sigma$ is
approximately constant over a wide range of radii
\citep[e.g.][]{Gerhard2001,Auger2010}.

The correlations between these characteristic measures of size, mass, and velocity
are typically considered separately for ellipticals and spirals, although this is starting to change with the advent of
integral field unit surveys, which have encouraged the use of more
sophisticated kinematic measures
\citep{deZeeuw2002,Cappellari2006,Weiner2006,Kassin2007,Cortese2014,Barat2019,Aquino2020}. Still,
considering E and S galaxies separately seems justified \citep[see e.g.][]{Ouellette2017}, not only because their velocity metrics and
morphologies differ, but also because they occupy rather distinct
regions in mass-velocity-size space.

Ellipticals are physically smaller than spirals at a given stellar mass
\citep{Shen2003}, and their population extends to higher masses than
spirals; most galaxies with $M_*>10^{11}\, \rm M_\odot$ are early-type,
while spirals are more prevalent in less massive galaxies
\citep[e.g.][]{Simard2011,Oh2020}. Their scaling laws also differ in
character -- for example, the rotation speeds of spirals scale tightly
with $M_*$ \citep[ the `Tully-Fisher relation', TFR;][]{Tully1977} --
but are nearly independent of galaxy size or surface brightness at given $M_*$
\citep{Zwaan1995,Courteau1999}.

On the other hand, the velocity
dispersion of ellipticals depends on both mass and size, roughly
tracing a 2D surface in the $M_*$-$\sigma$-$R_{\rm e}$ space
\citep[the `Fundamental Plane', FP;][]{Djorgovski1987,Dressler1987}. Despite these differences, the
scaling laws of both E and S galaxies are quite tight; indeed, both
the TFR and appropriate projections of the FP exhibit small
enough scatter to be used profitably as secondary distance indicators
\citep[see e.g.][and references therein]{Gallazzi2006,Pizagno2007}.

These qualitative differences between E and S galaxy scaling laws have
led a number of studies to adopt different frameworks to explain and
interpret the origin of these galaxies. The TFR, for example, has often been viewed in
a cosmological context as reflecting the deeper potential wells of
systems of increasing mass \citep[e.g.][]{Steinmetz1999,Bullock2001}. This is a
well-understood feature of cosmological models such as Lambda Cold
Dark Matter ($\Lambda$CDM), where the nearly scale-invariant power spectrum of
primordial mass fluctuations is modified during the
radiation-dominated era. Fluctuations that enter the horizon in that
era effectively stop growing because of the immense pressure exerted
by the coupled photon-baryon fluid. Since low-mass fluctuations enter
the horizon earlier their growth is suppressed relative to their more
massive counterparts, leading to systems where the potential deepens
(i.e. the `escape velocity' increases) with increasing mass
\citep[see e.g.][]{Mo2010}. If the
characteristic velocity of a galaxy somehow reflects the escape
velocity of its surrounding halo, then a tight scaling between
velocity and mass is naturally expected.

\begin{figure*}
\centering
  \includegraphics[width=0.8\linewidth]{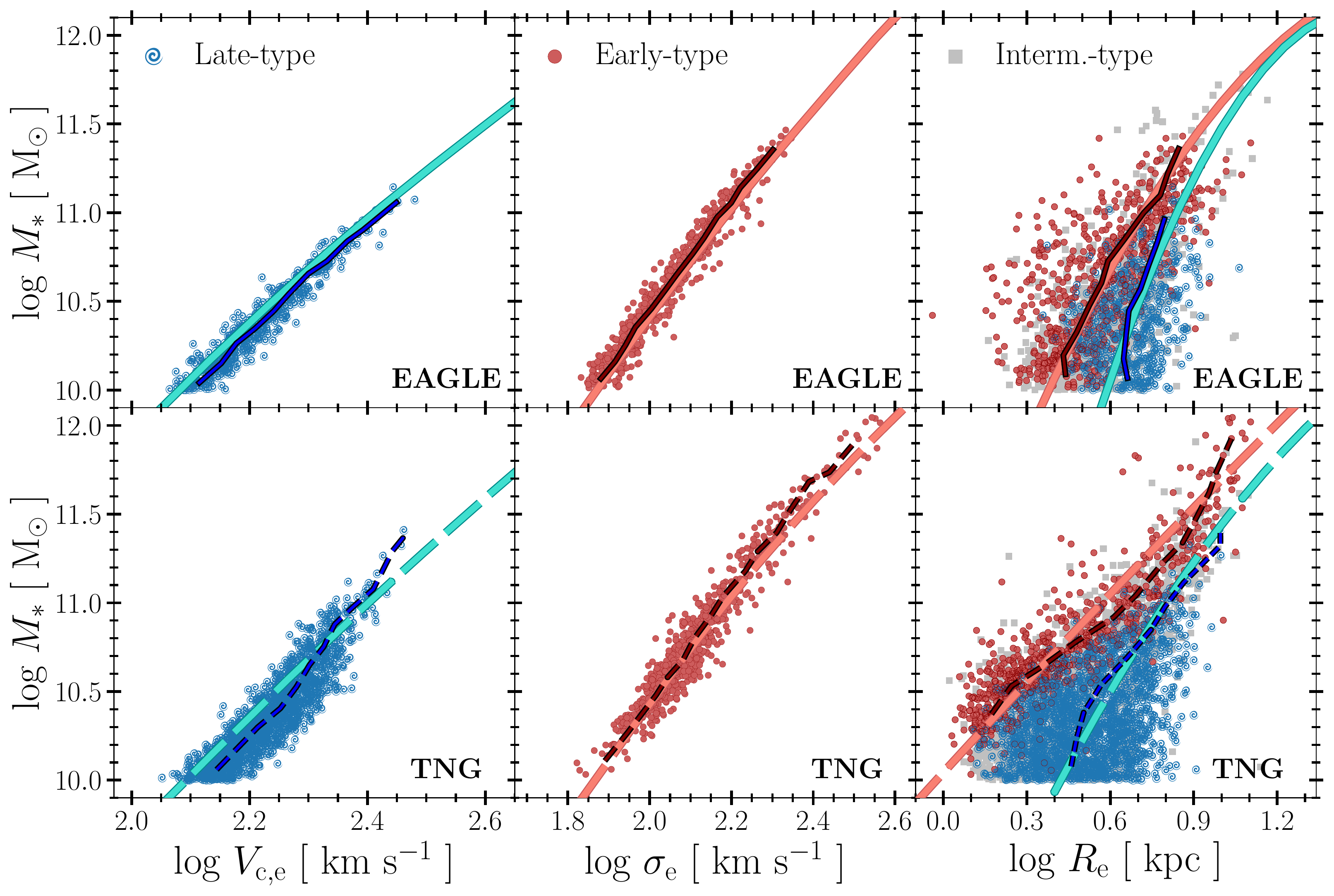}
  \caption{Correlations between the stellar mass, $M_*$, of simulated
    EAGLE (top row) and TNG (bottom row) galaxies and circular
    velocities at the stellar half-mass radius, $V_{\rm c,e}$, of
    late-type (S) systems (left column); line-of-sight velocity
    dispersion, averaged within the stellar effective radius of
    early-type (E) systems (middle column); and stellar half-mass
    radius (right column), for all galaxies in the sample. (See
    Sect.~\ref{SecMorphClass} for details on the morphological
    classification of simulated galaxies.) In all panels, red circles
    denote ellipticals, and blue spiral symbols denote spirals. Grey
    symbols are used to indicate other galaxies of intermediate
    type. Thick wiggly curves of matching colour indicate the median
    trends, computed in stellar mass bins. The thick smooth curves are
    fits used for (or obtained with) the fiducial model described in
    Sect.~\ref{SecVcReSims}.  }
    \label{FigScalingSims}
\end{figure*}

\begin{figure*}
    \includegraphics[width=0.8\linewidth]{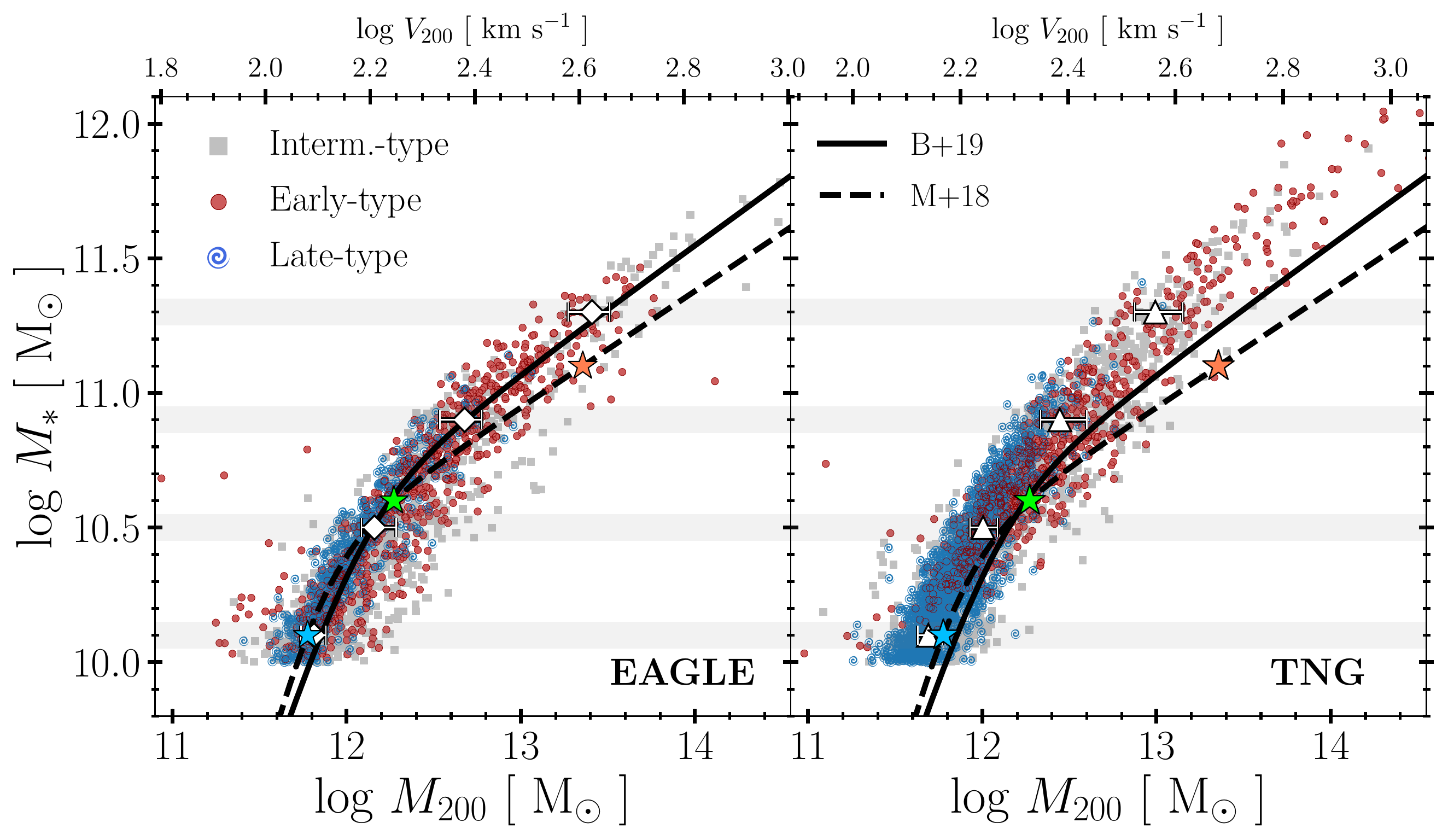}
\centering 
\caption{Galaxy stellar mass versus halo virial mass relation. Bottom
axes indicate $M_{200}$ in $M_\odot$; top axes indicate
  virial velocity, $V_{200}$, in km s$^{-1}$. The left-hand panel corresponds to
  EAGLE, right-hand panel to TNG. Symbols show individual galaxies
  coloured by morphology: Red circles indicate early-type (E); blue
  spiral symbols denote late-type (S); and grey squares are used for all
  others. Solid and dashed thick black curves indicate, for reference,
  the results of abundance-matching models from \citet{Behroozi2019}
  and \citet{Moster2018}, respectively. Open symbols (diamonds for
  EAGLE, triangles for TNG) indicate median $M_{200}$ values in the
  four bins of $M_*$ highlighted by the grey horizontal bands (these
  are used for the model curves shown in
  Fig.~\ref{FigVcReSims}). `Error bars' in each symbol denotes the
  $25$-$75$ percentile range in each bin. Starred coloured symbols correspond to 
  halo masses used for the models discussed in Fig.~\ref{FigVcReObs}.}
    \label{FigMgalM200}
\end{figure*}

One manifestation of these ideas is the equivalence between halo mass
and circular velocity imposed by the finite age of the Universe \citep{Mo1998}. This age
imposes a timescale that leads to a simple scaling between
virial\footnote{Virial quantities are identified by a `200'
  subscript and measured at the virial radius, $r_{200}$, defined as
  the radius where the enclosed mean density is $200$ times the
  critical density of the Universe, 
  $\rho_{\rm crit}=3H^2(z)/8\pi G$.} mass and circular velocity:
\begin{equation}
\label{EqM200V200}
M_{200} = \frac{V_{200}^3}{10\, G\, H(z)},
\end{equation}
where $H(z)$ is the Hubble constant at redshift $z$, and $G$ is Newton's gravitational
constant. This power-law scaling is close enough to
the TFR to suggest an interpretation where stellar masses and rotation
velocities scale roughly in proportion to the virial masses and
circular velocities of the halos they inhabit \citep[][]{Steinmetz1999,Navarro2000,Courteau2007}.

Although this may seem in principle plausible, a
simple proportionality between the galaxy and halo mass seems inconsistent
with the different shapes of the galaxy and $\Lambda$CDM halo mass functions
\citep{Guo2010,Moster2018,Behroozi2019}, and Eq.~\ref{EqM200V200} 
predicts a stronger evolution for the TFR with redshift than observed
\citep{Conselice2005,Flores2006,Miller2011}. In addition, this idea
does not explain why the rotation velocities of spirals should be
independent of size or surface brightness.

For elliptical galaxies, on the other hand, the origin of the FP is
usually ascribed to some variation of the `virial\footnote{This use of the term `virial' differs from that used in a
  cosmological context (Eq.~\ref{EqM200V200}). We shall distinguish
  between the two by referring to masses inferred from
  $M \propto \sigma^2 R$ as dynamical or `virial', in quotation
  marks.}'  theorem (VT), an idea motivated by the fact that
power-law fits to the empirical relations usually yield exponents not
too different from the virial relation, where dynamical masses
scale as $M_{\rm dyn}\propto \sigma^2 R_{\rm e}$
\citep[e.g.][]{Faber1987,Bernardi2003,Cappellari2006,Taranu2015}. On
closer scrutiny, however, the best-fit relations are significantly
different from `virial', and the implied dynamical masses are not
simply proportional to $M_*$. This leads to larger `mass-to-light
ratios' in more massive ellipticals \citep[e.g.][]{Gallazzi2006}, or
a `tilt' of the FP that has been ascribed to either systematic
changes in dark matter content, or to deviations from structural
homology, or to changes in the stellar initial mass function
\citep[e.g.][]{Ciotti1996,Jorgensen1996,Graham1997,Trujillo2004,Zaritsky2006}.

From a theoretical perspective, the relation between the scaling laws
of spirals and ellipticals has been addressed using semi-analytic
models of galaxy formation \citep[see e.g.][and references
therein]{Dutton2011,Desmond2017}, but are now also within the reach of
$\Lambda$CDM cosmological hydrodynamical simulations. This is especially true
of simulations able to follow statistically significant volumes and to
resolve the inner regions of individual galaxies, where the structural
parameters of galaxies are measured. These conditions are well met by
the latest round of cosmological hydrodynamical simulations, such as
the IllustrisTNG and EAGLE projects
\citep{Schaye2015,Crain2015,Pillepich2018,Springel2018}.  Although the
subgrid physics in these simulations has been adjusted to account for
some basic properties of the galaxy population, such as the galaxy
stellar mass function, neither the morphology of simulated galaxies
nor their detailed structure are directly prescribed by these
adjustments. The ability to reproduce observed galaxy scaling laws may
therefore be regarded as a genuine success (or failure) of the
simulations.

Recently, \citet{Ferrero2017} used the EAGLE simulations to study the
origin of the TFR in simulated galaxies morphologically classified as
`spirals'. Their analysis shows convincingly that the TFR emerges as
a result of (i) the tight correspondence between stellar mass and halo
mass (the $M_*$-$M_{200}$ relation); (ii) the non-linear structure of
cold dark matter halos; and (iii) the typical sizes of spirals.
Successfully reproducing the TFR depends on the convergence of these
three factors: $\Lambda$CDM simulations where galaxies tightly follow  the
$M_*$-$M_{200}$ relation inferred from `abundance-matching' (AM)
arguments and have sizes comparable to observed discs have no
difficulty reproducing the observed TFR.

The success of $\Lambda$CDM simulations at reproducing the observed TFR is
also due in no small part to the characteristic mass profile of cold
dark matter halos, which closely follows  the Navarro-Frenk-White
formula \citep[hereafter NFW,][]{Navarro1996,Navarro1997}. NFW
profiles are characterized by rising circular velocity profiles
in the very inner regions and an extended outer region where the
circular velocity is approximately flat. If spirals actually form in
the rising part of their surrounding halos, the contribution of their
luminous component can compensate for the contribution of the dark matter
to yield approximately flat outer rotation curves, as well as
characteristic rotation velocities that are nearly independent of
galaxy radius or surface brightness.  The requirement that galaxies form
in the rising part of the halo circular velocity profile places strong
constraints on the concentration parameter of NFW halos: This is not a free parameter of the model, but is fully specified by the
$\Lambda$CDM cosmological parameters \citep[see e.g.][]{Ludlow2016}. The
good agreement between observed and simulated TFRs thus provides
strong support for $\Lambda$CDM in the highly non-linear regime of the inner
regions of galaxies \citep{Navarro2019}.

The \citet{Ferrero2017} analysis also makes clear why early simulation
work found it so difficult to reproduce the observed TFR
\citep{Steinmetz1999,Navarro2000,Scannapieco2012}: Those simulations
either adopted the wrong cosmology, failed to match the appropriate
$M_*$-$M_{200}$ relation, or failed to reproduce observed disc sizes
at given stellar mass, or all of the above.

Intriguingly, the same analysis indicated that the EAGLE
$M_*$-$M_{200}$ relation depends little on the morphological type of
the simulated galaxy (see the left-hand panel of their Fig.~3). In other words,
S and E galaxies of a given stellar mass inhabit, on
average, halos of similar virial mass. This hints at the possibility
of reconciling the scaling laws of both S and E galaxies
within a unified scenario where the main difference between types is
their dark matter content or the importance of the baryons in setting
the characteristic velocity of a galaxy.

We explore these ideas here using samples of simulated and observed
galaxies of all morphological types.  We are particularly interested
in exploring scenarios where the scaling laws of all galaxies,
regardless of morphology, may be explained by a simple unified
physical model. Scaling laws for simulated galaxies in EAGLE and IllustrisTNG
have already been the subject of a number of studies
\citep[e.g.][]{Lagos2018,Rosito2019,vandeSande2019,Lu2020}, but to
our knowledge there has not been to date a concerted attempt to unify
the scaling relations of galaxies of different morphologies using such
simulations.

\begin{figure*}
    \includegraphics[width=0.8\linewidth]{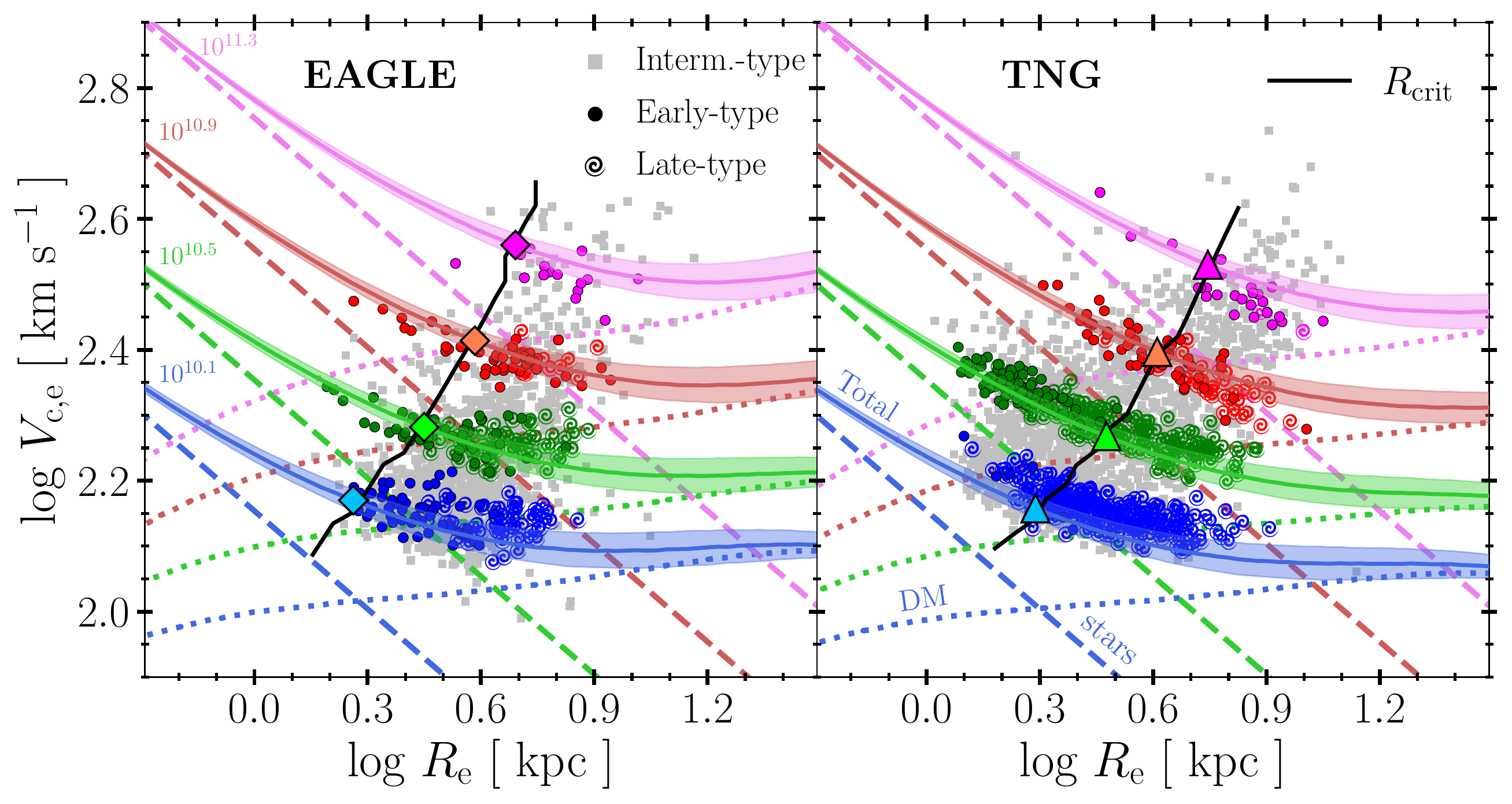}
\centering
\caption{Circular velocity at the stellar half-mass radius,
  $V_{\rm c,e}$, as a function of the effective radius, $R_{\rm e}$,
  for EAGLE (left-hand panel) and TNG (right-hand panel) simulated
  galaxies. All galaxies are shown in grey; those in four thin bins of
  stellar mass are highlighted in colour (see the grey bands in
  Fig.~\ref{FigMgalM200}). Median stellar masses for each bin are
  listed in the legend. Circles indicate early-type (E) galaxies,
  spirals denote late-type (S). Slanted dashed lines indicate the
  expected loci of galaxies that are fully dominated by their stellar
  component, that is, $V_{\rm c,e}^2=GM_*/2r_{\rm e}$. Dotted curves
  indicate the dark matter contribution expected for NFW halos with
  virial masses taken from the $M_*$-$M_{200}$ relation of each
  simulation (see, for example open symbols in Fig.~\ref{FigMgalM200}), contracted
  following the procedure of \citet{Gnedin2004}. Coloured lines
  correspond to adding in quadrature the dashed and dotted lines;
  error bands correspond to varying the NFW concentration about the
  average by $\pm 0.09$ dex. The coloured symbols (diamonds for
  EAGLE, triangles for TNG) indicate the `critical radius' for each
  stellar mass bin, defined as the galaxy radius for which the enclosed
  mass within $r_{\rm e}$ is split equally between dark matter and
  stars. See further discussion in the text.}
    \label{FigVcReSims}
\end{figure*}

\begin{figure}
    \includegraphics[width=\linewidth]{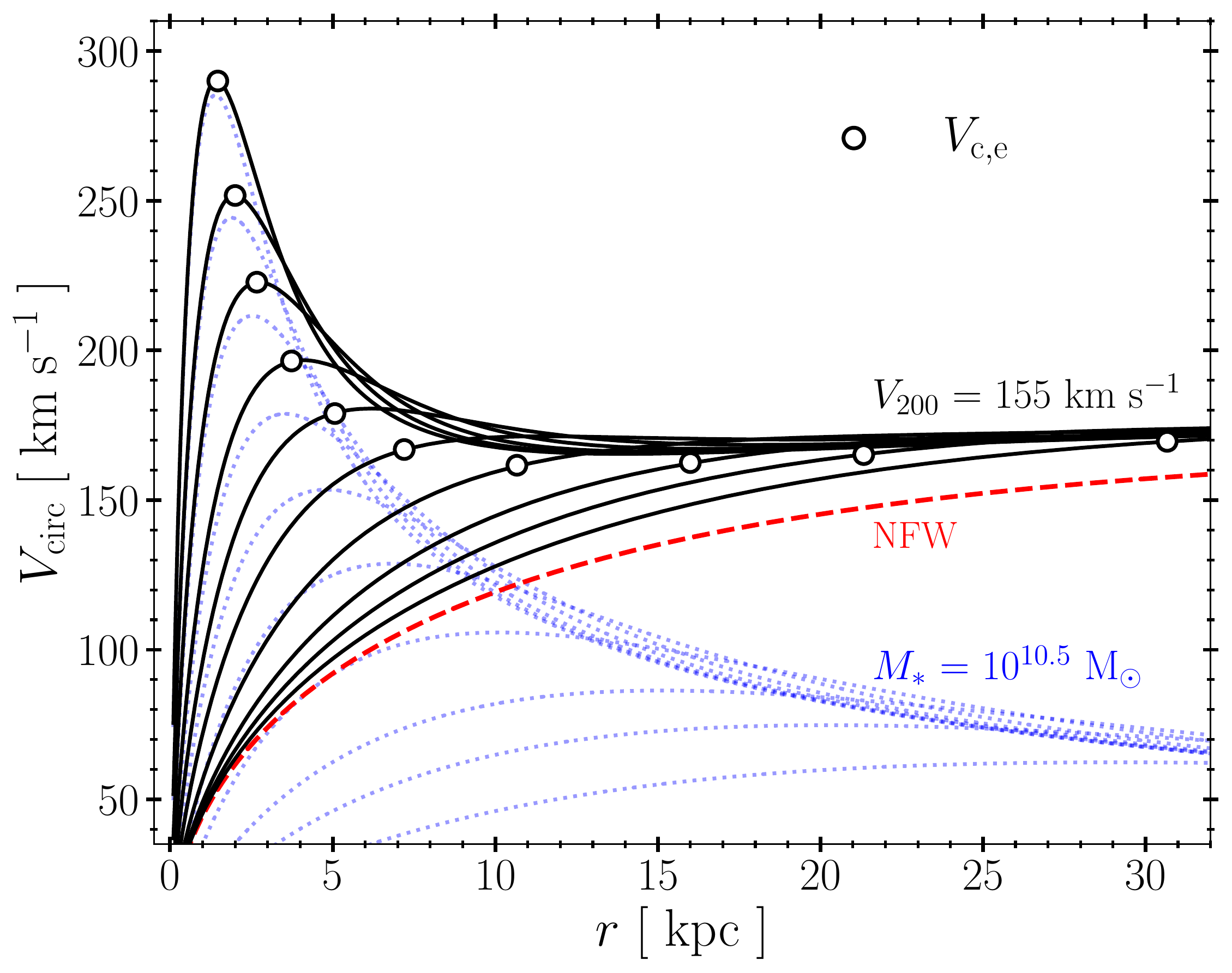}
\centering
\caption{Schematic circular velocity profiles (black solid lines) of
  galaxies of stellar mass $M_*=3\times 10^{10}\, \rm M_\odot$, embedded
  in an NFW halo (red dashed curve) of virial mass
  $M_{200}=1.4\times 10^{12}\, \rm M_\odot$ (or, equivalently,
  $V_{200}=155$ km s$^{-1}$) and average concentration, $c_{200}=7.6$
  \citep{Ludlow2016}. Different curves correspond to  exponential
  stellar discs of varying half-mass radius. The
  circular velocity of the dark matter plus stars at each of the
  half-mass radii, $V_{\rm c,e}$, is shown by the open circles. It is clear
  that $V_{\rm c,e}$ is approximately independent of $r_{\rm e}$ for
  galaxy radii exceeding a `critical' radius ($r_{\rm crit}$) of
  order $\sim 5$ kpc.  The characteristic rotation speeds of disc galaxies of given
  mass are thus roughly independent of radius, provided galaxy radii
  satisfy $r_{\rm e}>r_{\rm crit}$. Conversely, galaxies with
  $r_{\rm e}<r_{\rm crit}$ are expected to be dominated by their stellar
  component, and their characteristic velocities should depend
  sensitively on $r_{\rm e}$. }
    \label{FigModelRVc}
\end{figure}

It is important to note, before we begin, some limitations of this
study. One is that we focus here mainly on `luminous' galaxies, namely, those with stellar mass exceeding $\sim 10^{10}\, \rm M_\odot$,
for this is the regime that has been most aptly explored by $\Lambda$CDM
cosmological hydrodynamical simulations of large representative
volumes. It is also the regime where (cold) gas makes a relatively
small contribution, so we can use the stellar mass to approximate the
total baryonic mass of a galaxy. In addition, our study focusses only
on scaling laws linking mass, size, and velocity; there are, of
course, many other relations that may also be considered `scaling
laws', such as those linking star formation rates, gas mass
fractions, stellar ages and metallicities, supermassive black hole
mass, etc, but we shall not consider them here. Finally, we only
consider galaxies at $z=0$, although extending the ideas we explore
here to higher redshifts should be relatively straightforward, and we
plan to do so in future contributions.

The plan for this paper is as follows. Section~\ref{SecAnal} describes
the galaxy datasets we use, including our simulated and observed
galaxy samples, as well as the analysis procedure. Section~\ref{SecResSims}
presents and discusses the main results of our analysis of the
simulations while Section~\ref{SecResObs} discusses their application to
observed galaxies. We conclude with a brief summary in 
Section~\ref{SecConc}.

\begin{figure*}
    \includegraphics[width=0.9\linewidth]{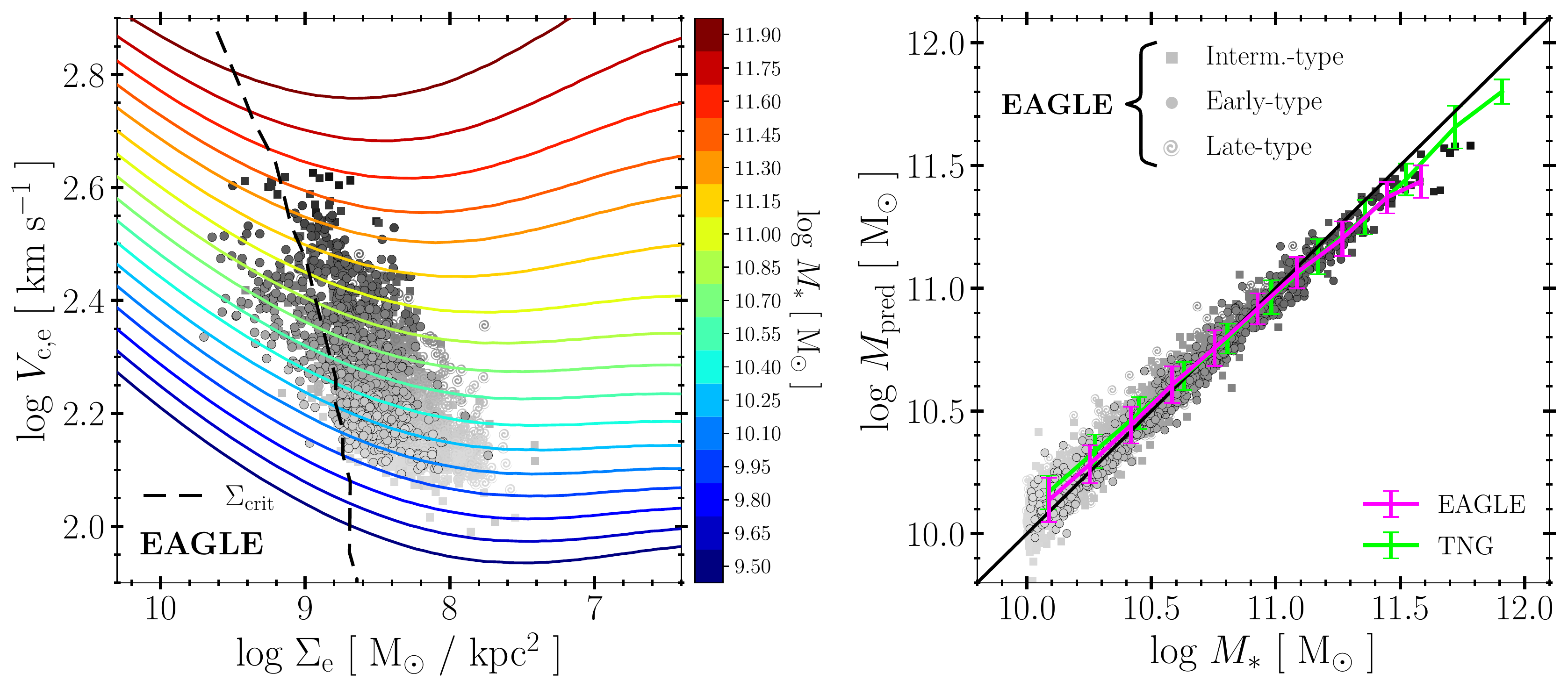}
\centering
\caption{Procedure to use the model as a secondary distance
indicator. {\it Left:} Grid of constant $M_*$ model curves linking
  $V_{\rm c,e}$ and the mean stellar surface density (brightness)
  within the effective radius,
  $\Sigma_{\rm e}\equiv M_*/2\pi R_{\rm e}^2$ (axis are inverted). Each curve is computed
  using the fiducial model described in Sect.~\ref{SecVcReSims}, using
  the EAGLE $M_*$-$M_{200}$ relation (see the left-hand panel of
  Fig.~\ref{FigMgalM200}). Galaxy stellar masses may be interpolated
  using this grid for all galaxies, regardless of morphological
  type. All EAGLE galaxies are shown shaded in grey according to
  stellar mass; circles for early-types, spirals for late types, and
  squares for the rest.  {\it Right:} Galaxy stellar masses predicted
  from the location of simulated galaxies in the
  $V_{\rm c,e}$-$\Sigma_{\rm e}$ plane on the left as a function of
  their true stellar mass. Is remarkable the excellent agreement and the rather
  small scatter around the median trend, shown by the magenta solid line
  (error bars indicate $25$-$75$ percentile range). Results of the
  same fiducial model, applied to the TNG dataset are shown by the
  thick green line (data for individual TNG galaxies not shown, for
  clarity). The good agreement suggests that the location of a galaxy
  in the $V_{\rm c,e}$-$\Sigma_{\rm e}$ plane can be effectively used
  as a powerful secondary distance indicator applicable to galaxies of
  all morphological types.}
    \label{FigMstrMpredSims}
\end{figure*}

\section{Data and analysis}
\label{SecAnal} 

We describe in this section the various datasets of simulated and
observed galaxies selected for this study. Regarding simulations, we
have chosen to analyse two $\Lambda$CDM cosmological hydrodynamical
simulations that have been shown to provide an adequate description of the
main properties of the galaxy population, such as their clustering and
stellar mass function at $z=0$. These simulations cover 
cosmological volumes large enough to yield representative galaxy samples and have
adequate spatial and mass resolution to enable kinematic measurements
at the half-mass radius of the stellar component, one of the basic
parameters used here.

Regarding observations, rather than relying on large catalogues such as
those made available by surveys like the Sloan Digital Sky Survey
(SDSS), we have chosen to base our study on smaller but highly curated
datasets that span a wide range of morphological types and that are
characterized by accurate photometry and detailed, spatially resolved
kinematic analysis. We note that these datasets are growing at steady
rate, mainly as a result of integral-field-unit (IFU) recent surveys
such as CALIFA \citep{Sanchez2012}, MANGA \citep{Bundy2015}, and SAMI
\citep{Scott2018}, among others. Those datasets have not
been incorporated in our analysis, but we are planning to
do so in future extensions of this work.

\subsection{Characteristic parameters}
\label{SecNotation}

Our analysis relies on three structural parameters per galaxy: their stellar
mass, $M_*$, as well as characteristic radii and velocities.  For
characteristic radii we shall use the stellar projected `effective' radius, 
$R_{\rm e}$, defined as the radius that contains, in projection,
half of the stellar mass, $M_*$. For ease of comparison with
simulations, we shall often translate this measure into a 3D stellar
half-mass radii, defined, for spheroidal galaxies,  by $r_{\rm e}=4 R_{\rm e}/3$. (Capitalized
`$R$' refers to projected radii, lowercase `$r$' to 3D radii.)

In terms of characteristic velocities, we shall use for simulated
galaxies the circular velocity at $r_{\rm e}$, $V_{\rm c,e}$. This is
well approximated in observed spirals by the rotation speed in the
asymptotic `flat' regime. We note that observational studies do not
always quote rotation velocities at $r_{\rm e}$, but they do strive to
infer the asymptotic characteristic value. Since most spirals have
relatively `flat' rotation curves, we adopt them as representative
of $V_{\rm c,e}$ without further corrections.  The circular velocity
at $r_{\rm e}$ is, of course, a direct measure of the total mass
enclosed within that radius;
$V_{\rm c,e}^2=GM_{\rm tot}(<r_{\rm e})/r_{\rm e}$ (assuming spherical
symmetry), which is readily available for all simulated galaxies.

For ellipticals, we shall use as characteristic velocity the stellar
line-of-sight velocity dispersion, $\sigma_{\rm e}$, averaged within
$R_{\rm e}$ and computed by averaging over three orthogonal
projections in the simulations. For observations, this is not a
directly measurable quantity, but it may be inferred from detailed
kinematic analysis of 2D velocity fields such as the ones available for
the observational datasets selected for this study.

\subsection{Simulations}
\label{SecSims}

\subsubsection{EAGLE}
\label{SecEAGLE}
The EAGLE\footnote{http://icc.dur.ac.uk/Eagle} project is a suite of
$\Lambda$CDM cosmological hydrodynamical simulations with parameters
consistent with results from \citet{PlanckCollaboration2014}: $\Omega_b = 0.0482$, $\Omega_{\rm dm} = 0.2588$,
$\Omega_{\Lambda} = 0.693$ and $h = 0.6777$, where
$H_0 = 100 \; h\ $km s$^{-1}$ Mpc$^{-1}$ is the present-day value of
Hubble's constant.  We refer the interested reader to
\citet{Schaye2015}, \citet{Crain2015} and \citet{McAlpine2016} for further details, and highlight here
only the basic features of the EAGLE run used for this work.

The simulation used here is Ref-L100N1504 \citep[see Table 1
of][]{Schaye2015}, which follows the evolution of $2 \times 1504^3$
particles in a periodic cubic volume of $100$ Mpc on a side from
redshift $z=20$ to $z=0$.  An equal number of dark matter and gas
particles are followed with a dark matter particle mass
$m_{\rm dm}=9.70 \times 10^6\; \rm M_{\odot}$ and initial gas particle
mass of $m_{\rm gas}=1.81 \times 10^6\; \rm M_{\odot}$. Gravitational
interactions are softened with a Plummer-equivalent scalelength of
$\epsilon = 2.66 $ kpc (comoving units) before redshift $z=2.8$ and
fixed at $\epsilon=0.7$ kpc (physical units) after that.

\subsubsection{IllustrisTNG}
\label{SecTNG}

We use also The Next Generation Illustris
Simulations\footnote{https://www.tng-project.org/} (IllustrisTNG)
\citep{Pillepich2018,Springel2018}, a suite of $\Lambda$CDM
magneto-hydrodynamic cosmological galaxy formation simulations. The
simulations include similar physical processes as EAGLE, but
implements them differently. In particular, the hydrodynamics is
simulated with the moving-mesh code AREPO \citep{Springel2010}, and
there are also subgrid physics details that may be consulted in the
above references. In our study here we use the run of a cubic box of
110.7 Mpc side length (TNG100; hereafter shortened to `TNG'), which
has been made publicly available \citep{Nelson2019}. The dark mass
resolution of the TNG100-full physics is
$m_{\rm dm} = 7.5 \times 10^6\, \rm M_\odot$, with a
Plummer-equivalent softening length of $0.74$ kpc.  The equivalent gas
(baryonic) mass resolution is
$m_{\rm gas} = 1.4 \times 10^6\, \rm M_\odot$. Gas cells are resolved
in a fully adaptive manner with a minimum softening length of $0.19$
kpc (comoving).

\subsubsection{Galaxy identification}
\label{SecIden}

Simulated galaxies are identified in both simulations using {\tt SUBFIND} \citep{Springel2001,Dolag2009}, a groupfinder that
  identifies self-bound `subhalos' in friends-of-friends (FoF) group
  catalogues constructed using a linking length of $0.2\times$ the mean
  interparticle separation. Each FoF has a `main' or `central'
  subhalo. The latter are the only ones we retain for our analysis here. The centre of each subhalo is chosen as the position of the
  particle with the minimum potential energy.

  Halo virial quantities are measured about that centre, including a
  virial radius, $r_{200}$, virial mass, $M_{200}$, and its corresponding
  circular velocity, $V_{200}$. Galaxy properties are defined using
  all particles inside a `galactic radius', defined as
  $r_{\rm gal}=0.15 \, r_{200}$. This definition ensures that the
  great majority of the stars associated with the central subhalo are
  included in the analysis. The projected stellar half-mass radius
  $R_{\rm e}$, and the inner line-of-sight velocity 
  dispersion, $\sigma_{\rm e}$, are computed averaging three
  orthogonal projections of each galaxy.

  Total stellar masses, $M_*$ are computed using all stars inside
  $r_{\rm gal}$. The circular velocity within the stellar half-mass
  radius is computed directly from the simulation data. Both EAGLE and
  TNG assume a \citet{Chabrier2003} stellar initial mass function
  (IMF). We focus here on `central' galaxies with a minimum stellar
  mass of $M_{*}=10^{10}\, \rm M_{\odot}$ (i.e. about $7\,000$ `star
  particles').

\subsubsection{Galaxy morphological classification}
\label{SecMorph}

Simulated galaxies may be assigned morphological types based on a
number of attributes. We choose here to categorize galaxies as
early-type (`elliptical') and late-type (`spiral') galaxies based
solely on two simple parameters: (i) the rotational-to-total kinetic
energy ratio parameter $\kappa_{\rm rot} = \sum V_{\rm xy}^2/\sum V^2 > 0.6$
\citep[][here $V$ is the magnitude of the total velocity
vector and $V_{\rm xy} \equiv j_z/R$ its azimuthal component perpendicular to the
$z$-direction, which is defined by the total angular momentum of the
galaxy's stellar component]{Sales2010} ; and (ii) the gas mass fraction within
$r_{\rm gal}$. These criteria are relatively strict as they
select only $22\%$ and $40\%$ as late types and $30\%$ and
$20\%$ as early types in EAGLE and TNG, respectively, at $z=0$. The
remaining are intermediate types, which we also consider in the
analysis. (See further details in Sect.~\ref{SecMorphClass} and
Fig.~\ref{FigMorphClass}.) Our final galaxy samples contain $2\,190$ EAGLE
galaxies ($483$ of them discs and $648$ ellipticals) and $3\,815$ TNG
galaxies ($1\,553$ of them discs and $731$ ellipticals).

\subsection{Observations}
\label{SecObs}

\subsubsection{Elliptical dataset: \rm{ATLAS}$^{\rm 3D}$}
\label{SecA3D}

We use in this work a sample of $258$ E galaxies from the
volume-limited ATLAS$^{\rm 3D}$ sample
\citep{Cappellari2013a,Cappellari2013b}.  Here, the effective radius
is defined as $R_{\rm e} = \sqrt{ {\rm A}_{\rm e} / \pi } $ where
$A_{\rm e}$ is the area of the effective isophote containing half of
the analytic total light of the Multi-Gaussian Expansion (MGE)
models. Velocity dispersions are measured by co-adding all spectra
contained within the "effective" ellipse with area
$A_{\rm e} = \pi {\rm R}_{\rm e}^{2} $. We use those velocities as
measures of $\sigma_{\rm e}$ without further correction. Finally, galaxy stellar masses are calculated from the
analytic total luminosity of the MGE model in the SDSS r-band and a
mass-to-light ratio of the stellar population within $R_{\rm e}$,
assuming a Salpeter IMF. In order to be consistent with simulated
data, we reduced stellar masses by $0.15$ dex in order to convert
from Salpeter to Chabrier IMF.

\subsubsection{Elliptical dataset: SLACS}

The Sloan Lens ACS (SLACS) Survey \citep{Bolton_2006} presents data
for a number of massive E galaxies selected largely on the basis of their
gravitational lensing power. \citet{Auger2010} present a subsample of 
59 galaxies confirmed as strong gravitational lenses
and with E or S0 morphologies. These are the galaxies we include
in our analysis. High-resolution multi-band Hubble Space 
Telescope (HST) imaging is used to infer stellar masses 
for each system using stellar populations synthesis (SPS) 
models and assuming a Chabrier IMF. Effective radii are determined
in each band and then used to infer the rest-frame $V$-band effective radius ($R_e$). Finally, 
velocity dispersions are measured within half of the effective radius. These velocities
are inferred from the luminosity-weighted stellar velocity dispersion within the 3" 
diameter aperture of the SDSS fibres and a theoretical prescription
\citep[see][for further details]{Auger2010}. We use those velocities as
measures of $\sigma_{\rm e}$ without further correction.

\subsubsection{Spiral dataset: SPARC}
\label{SecSPARC}
Part of our spiral dataset comes from the SPARC compilation
\citep{sparc2016}, a database of high-quality photometric and
kinematic data for 150 spiral and irregular galaxies.  Effective radii
are defined as those encompassing half of the total luminosity (K-band
or 3.6 $\mu$m).  \citet{sparc2016} assumes a stellar mass-to-light
ratio $\Gamma_{*}=0.5$ M$_{\odot}$/L$_{\odot}$ to convert luminosities
into stellar masses. We shall use their rotation velocity at the
effective radius as a measure of $V_{\rm c,e}$ without further
correction.

\subsubsection{Spiral dataset: Pizagno+07}
\label{SecP07}

Finally, the last observational sample we include in our analysis
corresponds to the catalogue of \citet[hereafter P+07]{Pizagno2007},
which contains a sample of 163 spiral galaxies with resolved
$H_{\alpha}$ rotation curves. Stellar masses are derived from the
luminosity using a constant I-band mass-to-light ratio of 1.2
\citep{Bell2003}. We assume that $V_{\rm c,e}$ corresponds to the
value of the rotation curves at $2.2 R_d$, where $R_d$ is the disc
exponential scale length.  Effective radii are taken to be equal to
the I-band half-light radius quoted in the catalogue.

\begin{figure*}
    \includegraphics[width=0.8\linewidth]{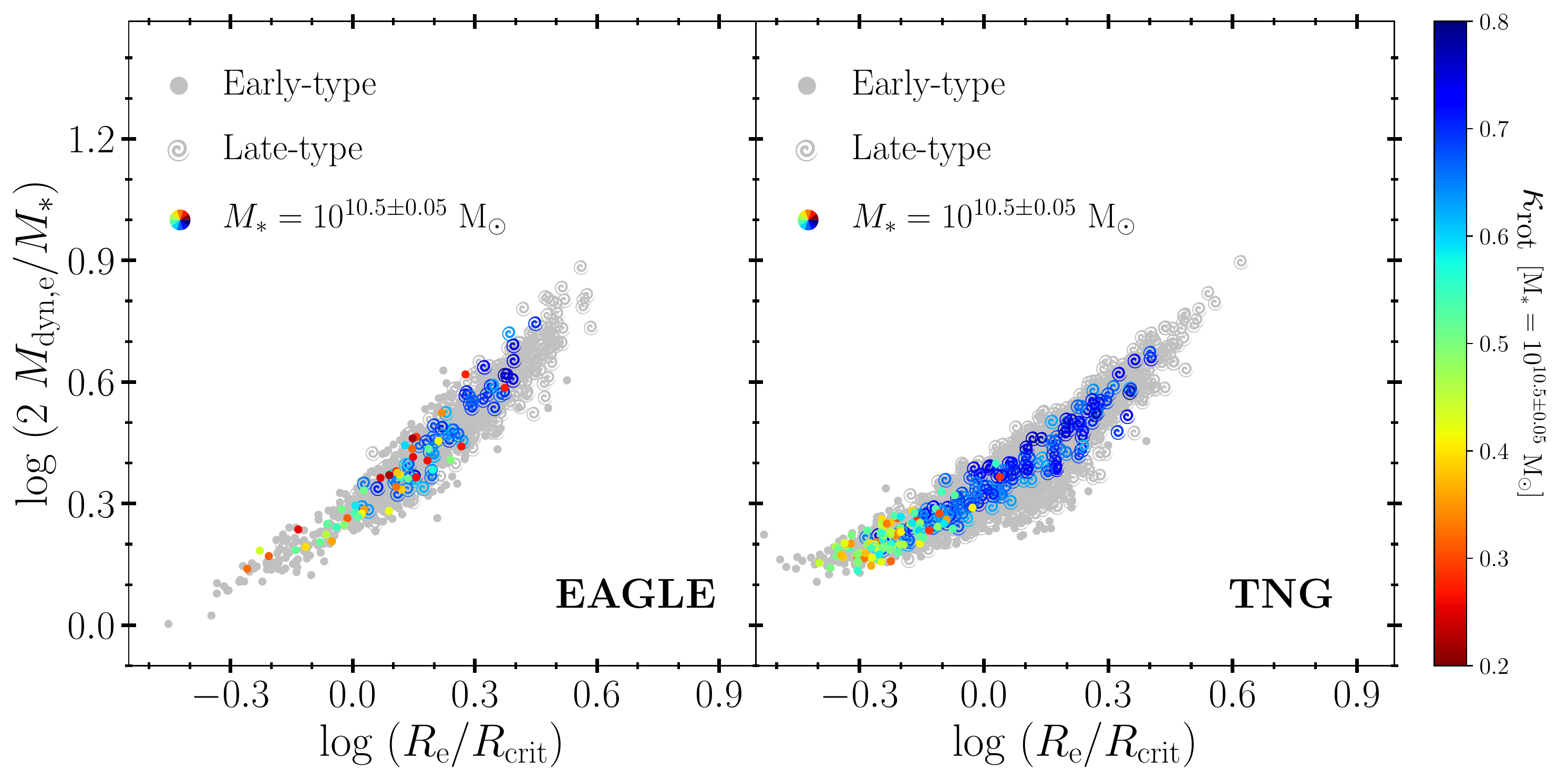}
\centering
\caption{`Dynamical mass-to-light ratio',
    $M_{\rm dyn,e}/(M_*/2)=V_{\rm c,e}^2\, r_{\rm e}/G(M_*/2)$, as a
    function of galaxy effective radius, expressed in units of the
    `critical' radius introduced in Fig.~\ref{FigModelRVc}, for
    EAGLE (left-hand panel) and TNG (right-hand panel).  Grey symbols represent
    all simulated galaxies, while those highlighted in colour
    correspond to galaxies with stellar masses
    $M_{*}=10^{10.5 \pm 0.05}$~M$_{\odot}$ (i.e. those in the second
    grey band, from bottom to top, in
    Fig.~\ref{FigMgalM200}). Galaxies in that mass bin are coloured
    according to $\kappa_{\rm rot}$, going from slow- (red) to
    fast-rotator galaxies (blue). It is clear that the dynamical
    mass-to-light ratio, which measures a galaxy's dark matter
    content, is an increasing function of $R_{\rm e}/R_{\rm
      crit}$. This suggests that galaxy size is a primary cause of the
    tilt of the FP.}
    \label{FigTiltSims}
\end{figure*}
\section{Simulation results}
\label{SecResSims}

\begin{figure*}
    \includegraphics[width=0.95\linewidth]{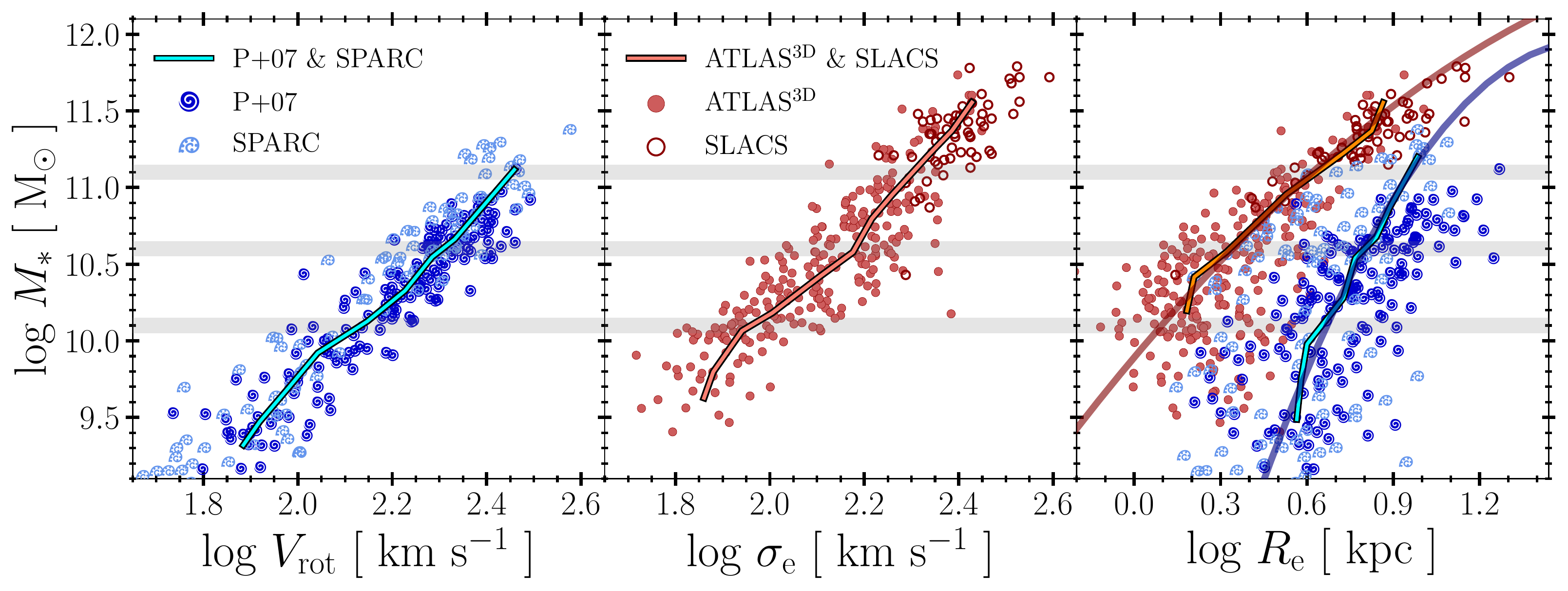}
\centering
\caption{Correlations between the stellar mass of galaxies in our
  observational sample and disc rotation
  speed at the stellar half-mass radius (disc galaxies
  from SPARC and P+07, left-hand panel); line-of-sight velocity
  dispersion, averaged within the stellar half-mass radius
  (ellipticals from ATLAS$^{\rm 3D}$ and SLACS; middle panel); and effective
  radius (all galaxies; right-hand panel). Circles denote ellipticals,
  spiral symbols denote discs. The thick wiggly curves indicate the median
  trend, computed in bins of stellar mass. The smooth coloured curves
  in the right-hand panel indicate fits to the average trends. }
    \label{FigScalingObs}
\end{figure*}

\subsection{EAGLE and TNG scaling laws}

We begin by exploring the scaling laws of simulated galaxies in
Fig.~\ref{FigScalingSims}. Here the left and middle columns show the
correlations between stellar mass and characteristic velocity for
systems identified as late-type (circular velocity, TFR) and
early-type (velocity dispersion). The latter is also known as the
`Faber-Jackson relation' \citep[hereafter, `FJR', for
short,][]{Faber1976}. The thick coloured wiggly lines trace the median
velocities as a function of stellar mass. As may be seen from this
figure, simulated galaxies show strong correlations between mass and
velocity; there is also good agreement between E and S galaxies in
EAGLE and TNG.  The main difference seems to be a somewhat increased
scatter in the TNG Tully-Fisher relation (bottom left-hand panel in
Fig.~\ref{FigScalingSims}) relative to EAGLE.

The right column in Fig.~\ref{FigScalingSims} shows the relation between
effective radius and stellar mass. (This figure also includes all
galaxies of intermediate galaxy types, shown in grey.)  Unlike the
correlations involving velocities, the trends between mass and size
are less clear and show much larger scatter; at fixed mass, galaxies
have effective radii that span nearly an order of magnitude. There
are also significant differences between the radii of EAGLE and TNG
galaxies. TNG galaxies span a wider range of radii and, particularly
at lower masses, the average radius of both E and S TNG galaxies appear
smaller than their EAGLE counterparts.

Why are the EAGLE and TNG velocity correlations in such good agreement
if their radii differ? Is the larger scatter in the TNG Tully-Fisher relation related to
the larger scatter of S galaxy radii? To answer these questions we need
to consider first whether TNG and EAGLE galaxies populate similar dark
matter halos.

\subsection{The galaxy stellar mass-halo mass relation}

This is done in Fig.~\ref{FigMgalM200}, where we show the galaxy
stellar mass versus virial mass relation for both EAGLE (left) and TNG
(right) galaxies. Open diamonds (EAGLE) and triangles (TNG) indicate
the median $M_{200}$ (as well as their respective $25$-$75$ percentile
range) for four thin bins of stellar mass. The `abundance-matching'
(AM) relations of \citet[][hereafter, B+19]{Behroozi2019} and
\citet[][hereafter, M+18]{Moster2018} are also shown, for
reference. These AM relations adopt the stellar
  mass functions of \cite{Bernardi2013,Bernardi2017}. Ellipticals
tend to inhabit, at given $M_*$, more massive halos, a trend that
agrees with what is inferred from observed satellite populations or
from galaxy lensing analyses \citep[see
e.g.][]{Wang2012,Mandelbaum2016}. The difference, however, is slight
and does not exceed a factor of $\sim 2$ in either
simulation. Ellipticals are also more prevalent at high masses while
spirals dominate at lower masses.

Fig.~\ref{FigMgalM200} also shows that, at fixed $M_*$, TNG halos have
systematically lower mass than EAGLE. The difference, however, is
small for $M_*<10^{11}\, \rm M_\odot$, with an offset in virial mass of
less than a factor of $\sim 0.25$ dex at fixed $M_*$. The difference is
more noticeable at higher masses, reaching a virial mass offset of a
factor of $\sim 4$ at $M_*\sim 3\times 10^{11}\, \rm M_\odot$. Despite
these differences, the FJR for massive EAGLE and TNG
ellipticals is nearly identical (middle column of
Fig.~\ref{FigScalingSims}). Also somewhat puzzlingly, the halo mass difference
between EAGLE and TNG for lower-mass galaxies does not seem enough to
account for the enhanced scatter of the TNG Tully-Fisher relation relative to EAGLE.

Since both E and S galaxies inhabit similar halos (at fixed $M_*$) it
is important to consider the role of galaxy radii in setting the
characteristic velocity of a galaxy. Indeed, both velocity dispersions (for E
galaxies) and rotation velocities (for S galaxies) are sensitive to
the total mass enclosed within the stellar half-mass radius,
$r_{\rm e}$, which we consider next.

\subsection{The circular velocity-size plane}
\label{SecVcReSims}

In order to analyse ellipticals and discs jointly, we consider for all
simulated galaxies the relation between their effective radii,
$R_{\rm e}$, and the circular velocity at the stellar half-mass
radius, $V_{\rm c,e}$. This is shown in Fig.~\ref{FigVcReSims}, where
the left-hand panel corresponds to EAGLE galaxies and the right-hand panel to
TNG. All galaxies are shown in grey but we highlight in colour those in
four $0.1$~dex-thin stellar mass bins centred at
$\log (M_*/\rm{M}_\odot)=10.1$, $10.5$, $10.9$, and $11.3$. (See the grey bands
in Fig.~\ref{FigMgalM200}.)  As in other figures, filled circles
denote early-type systems and spiral symbols indicate S galaxies.

Slanted dashed lines (labelled `stars') indicate the loci
expected for systems fully dominated by their stellar component, that is, 
$V_{\rm c,e}^2=G(M_*/2)/r_{\rm e}$, assuming spherical
symmetry. Dotted curves (labelled `DM') indicate the dark matter circular velocity
profile of NFW halos with virial mass taken from the $M_*$-$M_{200}$
relation of each simulation (see the open symbols in Fig.~\ref{FigMgalM200}). The NFW dark
halo profiles use concentrations from \citet{Ludlow2016} and are
scaled down, at all radii, by the universal baryon fraction, $M_{\rm dark}(r)=(1-f_{\rm bar})M_{\rm NFW}(r)$, with
$f_{\rm bar}=\Omega_{\rm bar}/\Omega_{\rm m}=0.186$, as appropriate for
a Planck-normalized $\Lambda$CDM cosmology. Here, $M_{\rm NFW}(r=R_{200})=
M_{200} $. The NFW profiles have also been
`contracted' to account for the effect of galaxy assembly, following
\citet{Gnedin2004}.

A simple fiducial model that predicts the loci of galaxies of
given stellar mass in this plane may be constructed by adding, in
quadrature, the dashed and dotted curves, respectively. This is shown
by the solid coloured curves in Fig.~\ref{FigVcReSims}, where the
`error bands' correspond to varying the NFW concentration of each
halo by the expected scatter of $0.09$ dex \citep{Ludlow2016}. It can be noted
that model curves of the same stellar mass differ between EAGLE and
TNG; this is due entirely to the slightly different $M_*$-$M_{200}$
relations of these two simulations (Fig.~\ref{FigMgalM200}). 

At fixed $M_*$, the circular velocity becomes nearly
  independent of galaxy size for galaxies with effective radii
  exceeding some `critical' radius, $R_{\rm crit}$. This threshold
  is reasonably well approximated by the solid black curve in each
  panel, which tracks, as a function of $M_*$, the effective radius of
  a galaxy expected, according to the fiducial model, to contain as
  much dark matter as stars within $r_e$ (i.e. the radii where the
  dotted and dashed lines intersect in Fig.~\ref{FigVcReSims}).

It may be somewhat surprising that ellipticals and spirals of given $M_*$
seem to follow approximately the same trends, despite the slight but
systematic differences in the $M_*$-$M_{200}$ relation of galaxies of
different morphology highlighted in Fig.~\ref{FigMgalM200}. This is
due to the fact that, for NFW models, the dark mass at fixed radii in the 
inner regions does not scale linearly with virial mass. Indeed, a
$M_{200}=10^{12}\, \rm{M}_\odot$ NFW halo of average concentration
encloses roughly $\sim 4.2\times 10^{9}\, \rm{M}_\odot$ within the
inner $3$ kpc. Varying the virial mass by a factor of two above and
below that value leads to variations of less than $20\%$ in the dark
mass enclosed within $3$ kpc. In other words, even relatively large
offsets in the  $M_*$-$M_{200}$ relation may yield nearly
imperceptible changes in the scaling relations.

Although we are encouraged by the relatively good agreement between
the fiducial model and simulation results, there are systematic
differences that are worth highlighting. For example, the model seems
to under-predict circular velocities for low-mass galaxies, and to
over-predict them for the most massive ones. These velocity offsets are clear,
but relatively small, typically of order $\sim 0.05$ dex, but suggest
that our fiducial model could be updated to improve
agreement. One possibility would be to consider different halo contraction
models. After all, there is evidence that massive ellipticals are
better modelled by an uncontracted halo \citep[see
e.g.][]{Shankar2017,Shajib2020}. We plan to consider such
corrections in future contributions.

\subsubsection{Reasons why is the Tully-Fisher relation independent of
  surface brightness}
\label{SecTFRSims}

Despite the simplicity of the fiducial model described above, it does
a remarkable job at reproducing the relation between characteristic
velocity and galaxy radius. The model shows that, at
fixed $M_*$, the characteristic velocity is expected to be roughly
independent of galaxy radius for a wide range of radii (i.e. at large
enough radii, curves of constant $M_*$ are nearly horizontal in
Fig.~\ref{FigVcReSims}). Indeed, only for galaxy radii smaller than the
`critical' value, $r_{\rm crit}$, the gravitational importance of
the stars raises the circular velocity of a system at $r_{\rm e}$ over
and above the asymptotic velocity of its surrounding halo \citep[see
e.g.][]{Mamon2005A,Mamon2005B}.

The reason for this is illustrated in Fig.~\ref{FigModelRVc}, where
the solid black curves show schematic circular velocity curves
corresponding to a fixed NFW halo (red dashed curve) and a galaxy of
fixed stellar mass but varying half-mass radius (shown by the open
circles). The galaxies are assumed to be exponential discs, and their
contributions to the circular velocity are shown by the blue dotted
curves. Because of the shape of the dark halo profile, the circular
velocity at the half-mass radius, $V_{\rm c,e}$ (shown by open
circles), barely changes for discs spanning a wide range of radii,
between $\sim 4$ and $\sim 30$ kpc. (For clarity, the halo, labelled
`NFW', is not `contracted' in this illustration.)

This is, of course, the main reason why the TFR is `independent of
surface brightness' in simulations \citep{Ferrero2017}: The TFR
depends solely on $M_*$, and reflects simply the scaling between $M_*$
and the characteristic velocity of its surrounding dark halo. The
series of models in Fig.~\ref{FigModelRVc} also make clear that this
independence should break down for galaxy radii smaller than some
`critical' radius, which we may define as the galaxy radius for
which the contribution to $V_{\rm c,e}$ of the dark and luminous
components is the same. This is the definitions used to compute
$R_{\rm crit}$ for each of the model curves shown in
Fig.~\ref{FigVcReSims} and show them as connected coloured diamonds
(EAGLE) or triangles (TNG). Clearly, the critical radius increases
with stellar mass, and that it is sensitive to the assumed
$M_*$-$M_{200}$ relation. Appendix~\ref{SecFits} provides an
analytical expression for the critical radius as a function of stellar
mass, which may be useful for the interested reader.

\subsubsection{The differences between TNG and EAGLE
  scaling laws}

The results of Fig.~\ref{FigVcReSims} may be used to explain the
trends and puzzles highlighted when comparing the TFR and FJR of EAGLE
and TNG in Fig.~\ref{FigScalingSims}. We consider first why the TFR in
EAGLE exhibits so little scatter, even though late-type galaxy radii
span a wide range. This is simply because S galaxies in EAGLE have
relatively large radii, which exceed in all cases their corresponding
values of $R_{\rm crit}$. As discussed above, in this regime the
characteristic velocity is independent of galaxy radius and depends
solely on the $M_*$-$M_{200}$ relation. The scatter in the EAGLE TFR
thus reflects the scatter in that relation, which is rather small
\citep[see][for a more detailed discussion]{Ferrero2017}.

The same argument explains why the TNG Tully-Fisher relation has enhanced scatter
compared to EAGLE (see the left-hand panels in Fig.~\ref{FigScalingSims}). This
is not because of enhanced scatter in the $M_*$-$M_{200}$ relation,
but rather because spiral galaxy radii in TNG cover a wider range than
in EAGLE, and include many systems with radii comparable to, or
smaller than, the critical radius for their $M_*$. These galaxies have
therefore more dominant stellar components and higher characteristic
velocities, as may be seen for systems with $r_{\rm e}<r_{\rm crit}$
in Fig.~\ref{FigModelRVc}. These `small' galaxies scatter off the
TFR towards higher velocities at fixed $M_*$, yielding increased
dispersion in the TNG Tully-Fisher relation relative to EAGLE.

One may also use these arguments to explain the invariance of the FJR
in EAGLE and TNG (middle column in Fig.~\ref{FigScalingSims}). This is
due to the rough compensation of two effects: TNG ellipticals are
smaller on average than EAGLE's, which pushes their velocities high,
but, at fixed $M_*$, they inhabit systematically lower-mass halos,
which pushes their velocities down. The combination of the two effects
results in FJR relations that are basically indistinguishable for TNG
and EAGLE. We shall return to this issue in more detail when comparing
with observations in Sect.~\ref{SecCompSimsObs}.

\subsection{A possible new distance indicator}
\label{SecDistIndSims}

A simple application of these ideas leads to a new secondary distance
indicator that may be applied indistinctly to S and E
galaxies, by exploiting the unique mapping between the location of a
galaxy in the $V_{\rm c,e}$-$R_{\rm e}$ plane and its stellar
mass. This mapping still holds if we substitute galaxy radii by
the mean stellar surface density of a galaxy,
$\Sigma_{\rm e}\equiv (M_*/2)/ \pi R_{\rm e}^2$, with the advantage
that both velocity and surface density/brightness are
independent of distance.

We show this in the left-hand panel of Fig.~\ref{FigMstrMpredSims}
where all EAGLE galaxies are shown in grey (shaded to indicate stellar
mass) and 
superposed on a grid of models akin to those shown in
Fig.~\ref{FigVcReSims}, but for a finer spacing in stellar mass. Each model curve denotes the loci of constant $M_*$, so that any
galaxy in this plane may be assigned a stellar mass interpolated from
this grid. 
Each curve results from varying $R_e$ in the fiducial
model at fixed $M_*$, assuming a galaxy mass-halo mass relation and
contracted halos of average concentration, as described in Sect.~\ref{SecVcReSims}.
We show the result of this interpolation in the right-hand
panel of Fig.~\ref{FigMstrMpredSims}, where we compare `predicted'
values of the stellar mass with `true' values measured from the
simulation. The agreement between the two is remarkable, especially
given the simplicity of the model, with little bias and an
exceptionally small scatter of only $0.078$ dex around the $1$:$1$ relation.

This suggests a secondary distance indicator, where the measurement of
a characteristic velocity and an effective surface brightness may be
combined to yield a prediction for the total stellar mass. Comparing
this to the apparent magnitude of a galaxy would yield its distance in
a straightforward way. This secondary distance indicator is, of
course, just a combination of (i) the TFR for `large' (typically
spiral) galaxies, which have $R_{\rm e}>R_{\rm crit}$ (or,
equivalently, $\Sigma_{\rm e} < \Sigma_{\rm crit}$) and for which the
characteristic velocity depends solely on stellar mass (i.e. curves
of constant $M_*$ are nearly horizontal in Fig.~\ref{FigMstrMpredSims}
for $\Sigma_{\rm e}<\Sigma_{\rm crit}$), and (ii) an FP-like relation for `small' (typically elliptical) galaxies with
$R_{\rm e}<R_{\rm crit}$ ($\Sigma_{\rm e} > \Sigma_{\rm crit}$) , for
which the characteristic velocity depends on both velocity and 
stellar surface density/brightness.

It should be noted that the only information used to construct the grid shown
in the left-hand panel of Fig.~\ref{FigMstrMpredSims} is the EAGLE
galaxy stellar mass-halo mass relation and the assumption that $\Lambda$CDM halos are
well represented by NFW profiles. The same procedure may be applied to
TNG, after modifying the model grid to account for the slightly different
$M_*$-$M_{200}$ relation. The results are also shown in the right-hand
panel of Fig.~\ref{FigMstrMpredSims}, with similar results (little
bias and a scatter of only $0.075$ dex). We shall return to these
issues when applying these results to observed galaxies in
Sect.~\ref{SecResObs}.

\begin{figure}
    \includegraphics[width=0.9\linewidth]{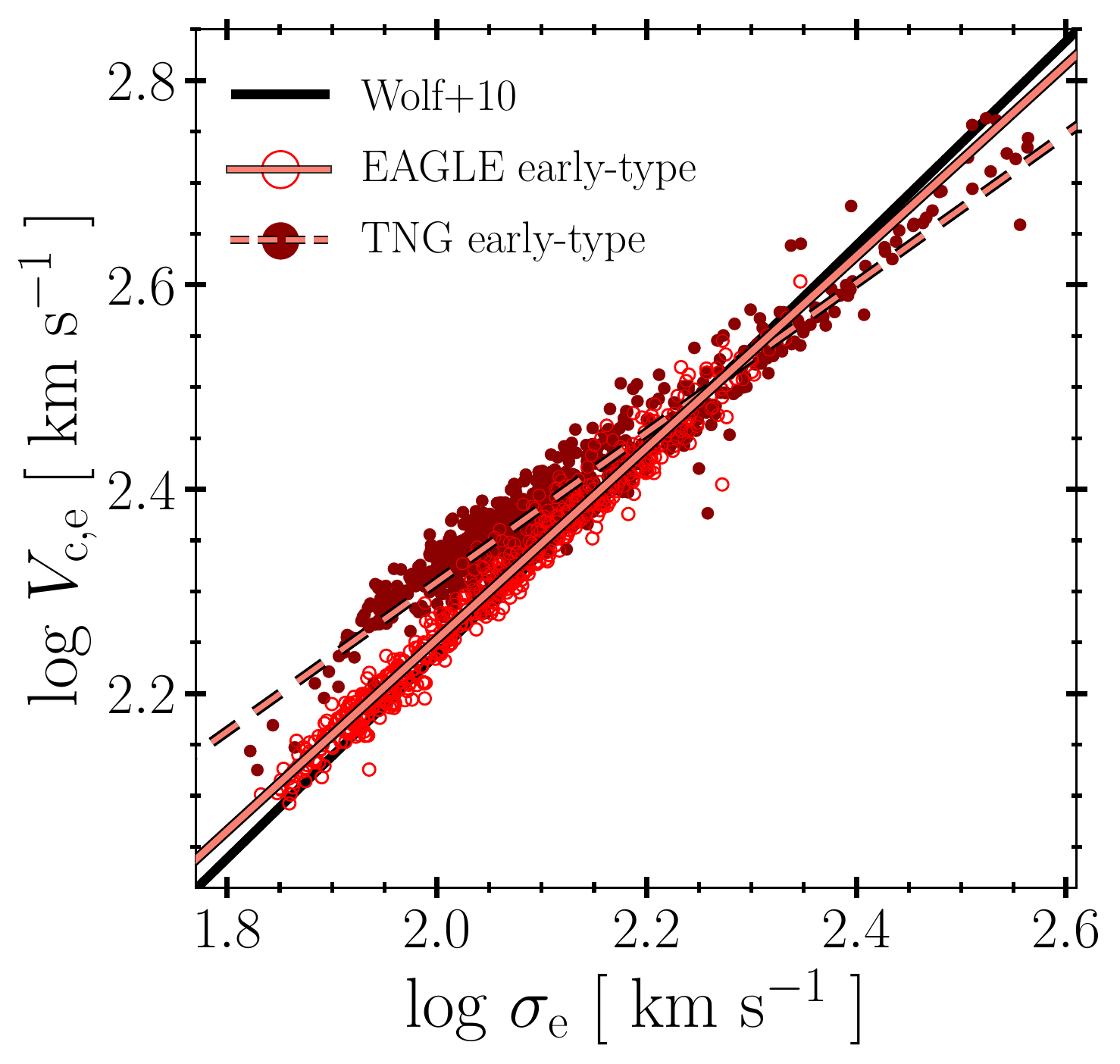}
\centering
\caption{Circular velocity at the stellar half-mass radius,
  $V_{\rm c,e}$ as a function of the stellar line-of-sight velocity
  dispersion, averaged within the effective radius, $R_{\rm e}$. All
  galaxies categorized as early-type in EAGLE (open circles) and TNG
  (filled circles) are shown. (See Sect.~\ref{SecMorphClass} for
  details on the morphological classification procedure.) The thick
  black solid line indicates the model of \citet{Wolf2010}. The solid
  and dashed red lines are simple power-law fits to the EAGLE and TNG
  data, respectively. For fits of the form
  $V_{\rm c,e}=V_0\, (\sigma_e/$km s$^{-1})^n$, the best fit parameters are
  $(V_0,n)=(2.37~$km s$^{-1},0.94)$ for EAGLE and
  $(V_0,n)=(7.03~$km s$^{-1},0.73)$ for TNG.}
    \label{FigVcSigma}
\end{figure}

\subsection{The `tilt' of the fundamental plane}
\label{SecTiltSims}

The analysis of Fig.~\ref{FigVcReSims} may also be applied to shed
light on the interpretation of a possible `tilt' to the FP. This is commonly used to refer to the fact that the best
fitting `plane' for elliptical galaxies in the stellar
mass-velocity-radius space differs from that expected from the `virial'
relation, $M_*\propto \sigma^2 R_{\rm e}$.

For this scaling to hold, stars must be gravitationally dominant over
the dark matter at $R_{\rm e}$. In the language of
Fig.~\ref{FigVcReSims}, this is equivalent to having galaxies closely
hug the dashed curves in this figure, or, equivalently, galaxies with
radii much smaller than `critical'. Although some such galaxies
exist, they are few; most TNG and EAGLE ellipticals straddle radii
around the critical value. Their velocities are therefore heavily
affected by the dark matter component, leading them to deviate from
the dashed lines and, consequently, from the `virial' scaling
\citep[see e.g.][for a similar discussion]{Zaritsky2006}.

In other words, the `tilt' in the simulated FP is
related to the fact that, at fixed $M_*$, physically larger galaxies
simply contain more dark matter than smaller ones. We show this in
Fig.~\ref{FigTiltSims} where we plot, for all S and E 
galaxies in TNG and EAGLE, the `dynamical mass-to-light ratio' as a 
function of galaxy radius, expressed in units of the `critical' value. 
This ratio is defined here as that between total mass and stellar
mass within $r_{\rm e}$, that is, 
$M_{\rm dyn,e}/(M_*/2) \equiv V_{\rm c,e}^2 r_{\rm e}/G(M_*/2)$.  A ratio
close to unity thus indicates that stars are fully dominant.

Expressed in this manner, we see in Fig.~\ref{FigTiltSims} that the
dynamical mass-to-light ratio increases monotonically with
$R_{\rm e}/R_{\rm crit}$, for all galaxies, regardless of stellar
mass. This insight may in principle be checked observationally as it
predicts that subsets of ellipticals that are `compact' (in the
sense of $R_{\rm e} \ll R_{\rm crit}$) should exhibit smaller tilt,
and also that the tilt should exist even for ellipticals of fixed
stellar mass, provided they span a wide enough range of galaxy
radii.

These results also suggest that rotational support is
  secondary to galaxy size in determining the magnitude of the
  tilt. Indeed, the coloured symbols in Fig.~\ref{FigTiltSims}
  indicate the importance of rotation (measured by $\kappa_{\rm rot}$)
  for galaxies with stellar masses
  $M_{*}=10^{10.5 \pm 0.05}$~M$_{\odot}$ (i.e. green symbols in
  Fig.~\ref{FigVcReSims}). For such galaxies, there is no clear-cut
  indication that rotation plays a major role at fixed galaxy size,
  although size and rotational support are clearly correlated at fixed
  $M_*$. Our results thus suggest that the `tilt' should be
  minimized when considering samples of fixed $R_e/R_{\rm
    crit}$. Choosing `slow' or `fast rotators' should in principle
  yield samples with systematically low or high $R_e/R_{\rm crit}$. This may
  explain the systematic differences in the FP of slow versus fast
  rotators reported by \citet{Bernardi2020}. We plan to address this
  issue more directly, and with an enhanced observational sample, in
  future work.

\begin{figure}
    \includegraphics[width=0.9\linewidth]{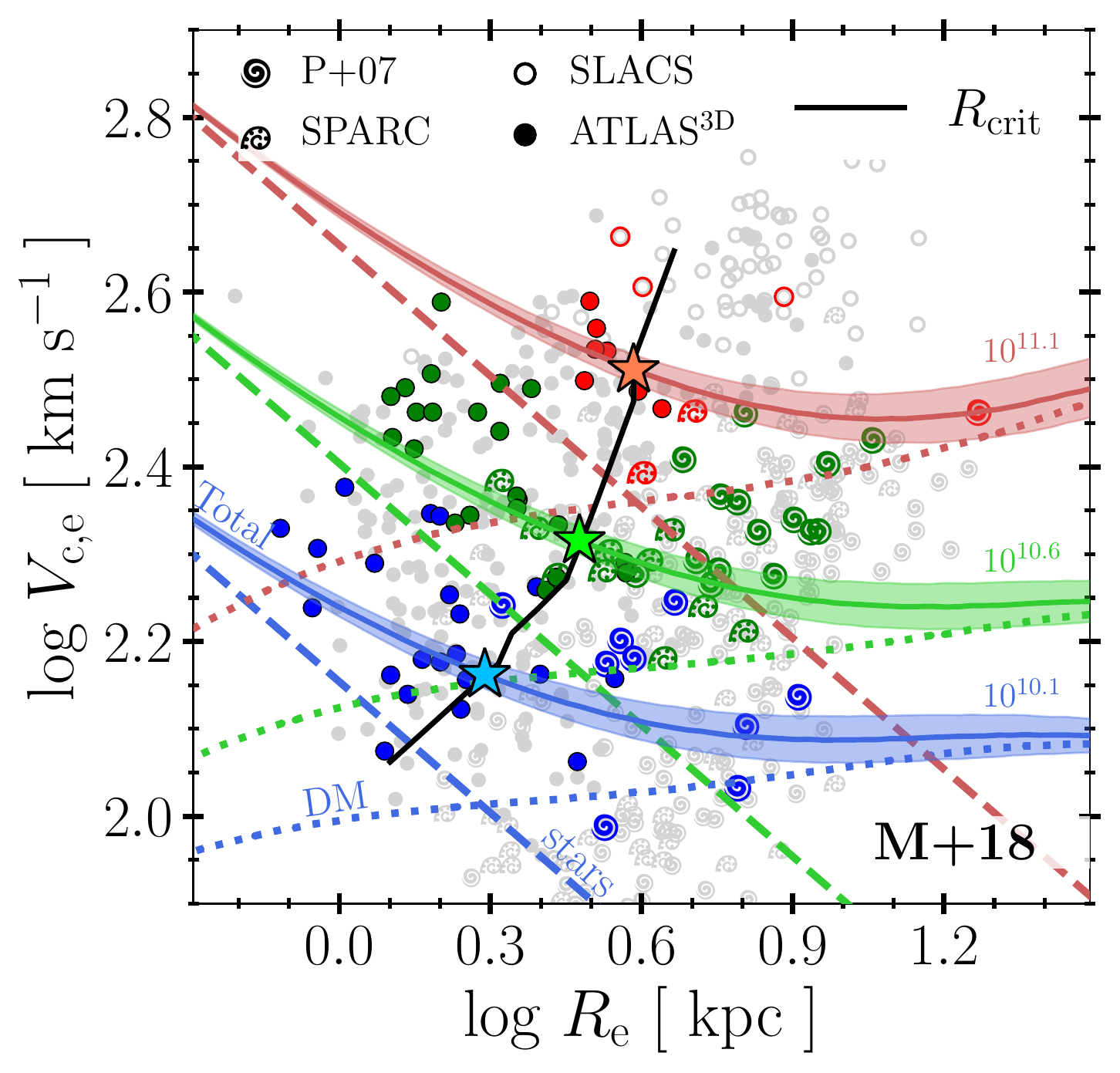}
    \centering
    \caption{Same as Fig.~\ref{FigVcReSims}, but for observed galaxies in
      our sample. Circular velocities, $V_{\rm c,e}$, are assumed to
      be equal to the characteristic rotation speed for galaxy discs
      (spiral symbols in Fig.~\ref{FigScalingObs}) and are inferred
      from the \citet{Wolf2010} relation for ellipticals
      (circles). Halo masses for each of the three stellar mass bins
      highlighted in coluor are chosen from the M+18 abundance-matching
      relation, as shown by the starred symbols in
      Fig.~\ref{FigMgalM200}. Coloured stars joined by a solid line
      indicate the `critical' radius where the dark matter and
      stellar mass within the stellar half-mass radius are equal.}
    \label{FigVcReObs}
\end{figure}

\begin{figure*}
    \includegraphics[width=0.9\linewidth]{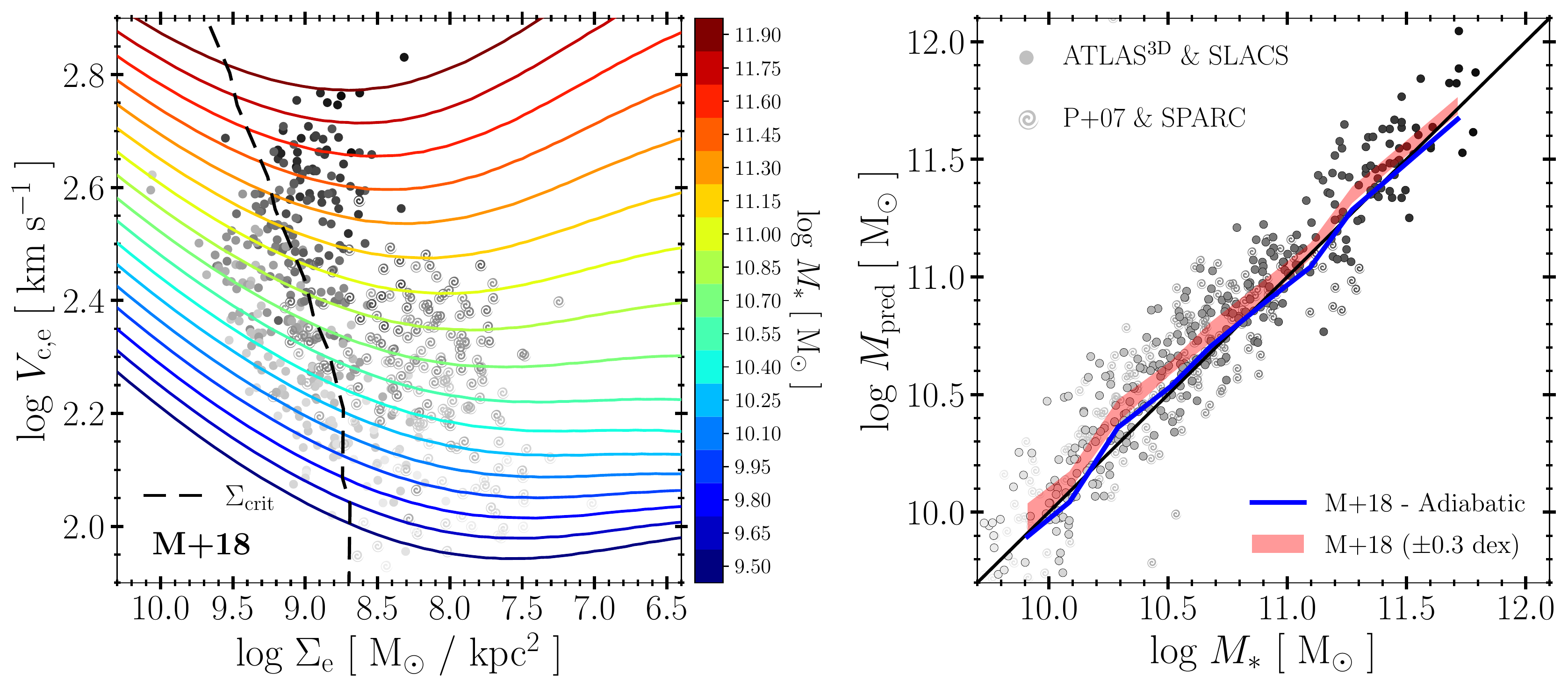}
    \centering
    \caption{Same as Fig.~\ref{FigMstrMpredSims}, but for our observed
      galaxy sample. The panel on the left shows the grid of stellar
      mass models computed using the fiducial model described in
      Sect.~\ref{SecDistIndSims}, assuming the M+18 abundance-matching
      $M_*$-$M_{200}$ relation. Line colours indicate stellar mass, as
      indicated in the colour bar. Observed galaxies are shown by
      symbols in shades of grey, according to their stellar
      mass. Circles are used for early-types; spiral symbols for late
      types. Interpolation on this model grid allows stellar masses to
      be `predicted' solely from the location of a galaxy in the
      $V_{\rm c,e}$-$\Sigma_{\rm e}$ plane. The predicted values of
      $M_*$ are shown as a function of the true values of $M_*$ in the
      right-hand panel. The median trend is shown by the thick pink
      line, with an `error band' that spans the change in the median
      predicted values after varying the $M_*$-$M_{200}$ relation by
      a factor of two in halo mass above and below the
      M+18 relation. The blue curve shows the median
      trend, but for a model that includes `adiabatic' contraction
      instead of the \citet{Gnedin2004} contraction model. }
    \label{FigMstrMpredObs}
\end{figure*}

\section{Application to observed galaxies}
\label{SecResObs}

\subsection{Scaling laws}

We now apply these ideas to the observed sample
(Sect.~\ref{SecObs}), which contains an assortment of galaxies of
different morphological types, ranging from massive ellipticals to
late-type spirals spanning a wide range of mass. Their scaling laws
are presented in Fig.~\ref{FigScalingObs}, where the left-hand panel shows
the TF relation of disc galaxies and the middle panel
shows the FJR for ellipticals. It should be noted that both are rather tight
relations, with a mass scatter of $0.28$ dex and $0.27$ dex,
respectively, about the median velocity trends traced by the thick
wiggly curves in each panel.

As for simulated galaxies, the mass-size relation
(right-hand panel in Fig.~\ref{FigScalingObs}) exhibits much larger
scatter, spanning nearly a decade in galaxy radius at fixed
$M_*$. Still, there are strong trends once the sample is parsed into
morphological types, and it is clear that the average size increases
with increasing mass for both early and late types.

It is also clear from comparing Fig.~\ref{FigScalingObs} and
Fig.~\ref{FigScalingSims} that there are strong similarities between the
scaling laws of observed and simulated galaxies. This motivates us to
carry out the same analysis as in the previous section, where, in
order to compare galaxies of different morphologies, we estimate their
total mass (or, equivalently, circular velocity) enclosed within
the stellar half-mass radius, $r_{\rm e}$. In the case of spirals,
assuming that the rotation speed at $r_{\rm e}$ traces the circular
velocity seems reasonable, in other words, 
$V_{\rm c,e}\equiv V_{\rm rot}$. For elliptical galaxies this is
less straightforward, but there is compelling literature arguing that
the line-of-sight velocity dispersion may also be used as a tracer of the
circular velocity at $r_{\rm e}$ \citep[see e.g.][and references
therein]{Walker2009,Wolf2010}.

We show this in Fig.~\ref{FigVcSigma} for all E galaxies in
EAGLE (open circles) and TNG (filled circles). Red solid and dashed
lines indicate power-law fits to the $\sigma_{\rm e}$-$V_{\rm c,e}$
relation, with circular velocity RMS of $0.05$ dex and $0.07$ dex,
respectively. The solid black line indicates the result of the model
of \citet{Wolf2010}, $V_{\rm c,e}=\sqrt{3}\, \sigma_{\rm e}$, which
provides a less accurate but acceptable fit to the combined data, with
an RMS of $0.12$ dex. For simplicity, we shall hereafter adopt this
relation in order to estimate $V_{\rm c,e}$ for observed ellipticals
in our sample, although one could in principle improve on this
prescription by using a more sophisticated analysis \citep[see
e.g.][for a review]{Courteau2014}.

\subsection{The circular velocity-size plane}

The assumption that $V_{\rm c,e}=\sqrt{3}\, \sigma_{\rm e}$ enables us
to place all observed galaxies in the $V_{\rm c,e}$ versus $R_{\rm e}$
plane, as shown in Fig.~\ref{FigVcReObs}. Following a similar analysis
to that adopted when discussing Fig.~\ref{FigVcReSims}, we highlight
in colour galaxies in three thin bins of stellar mass (the bins are
indicated by shaded grey bands in
Fig.~\ref{FigScalingObs}). Reassuringly, we see the same qualitative
behaviour as for simulated galaxies, albeit with increased
scatter. The trends are well reproduced by the same fiducial model
introduced in Sect.~\ref{SecVcReSims}, after assuming a halo virial mass for
each of the three bins.

For the three bins of stellar mass centred at
$\log{M_*/\rm{M}_\odot}=10.1$, $10.6$, and $11.1$, halo masses of
$\log{M_{200}/\rm{M}_\odot}=11.56$, $12.1$, and $13.28$ result in
adequate fits for both elliptical and spiral galaxies in each
bin. (As in Fig.~\ref{FigVcReSims} the coloured bands show the result
of the fiducial model, including a variation of $0.09$ dex in the NFW
concentration). These halo mass values are shown as starred symbols in
Fig.~\ref{FigMgalM200}, and have been chosen to lie on the M+18
abundance-matching relation. The agreement is encouraging, for it
implies that, as in the simulations, the data seem consistent with the
idea that there are, at fixed $M_*$, no major differences between the
velocity-radius relation of galaxies of different morphological types.


\subsection{The circular velocity-surface brightness plane as distance indicator}

We can take the modelling one step further, and, as in
Fig.~\ref{FigMstrMpredSims}, we can use `predict' stellar masses
based on the location of galaxies in the $V_{\rm
  c,e}$-$\Sigma_{\rm e}$ plane, together with an assumed
$M_*$-$M_{200}$ relation and a halo contraction model.  We show the
grid of `fiducial' models (see Sect.~\ref{SecVcReSims}) in the left-hand panel of
Fig.~\ref{FigMstrMpredObs} that result from assuming the M+18
abundance-matching relation. The `predicted versus true' stellar masses
are shown in the right-hand panel of the same figure, and show
encouraging results. The predicted masses show overall little bias
(predicted masses are on average just $\sim 0.1$ dex higher than
observed) and an RMS of only $0.19$ dex about the $1$:$1$ relation.

The agreement between predicted and observed stellar masses may be
improved by adopting a different $M_*$-$M_{200}$ relation, or by
refining the very simple model used to construct the grid shown in the
left-hand panel of Fig.~\ref{FigMstrMpredObs}. Interestingly, the
results are fairly insensitive to changes in the $M_*$-$M_{200}$
relation. Indeed, the pink `error band' shows the effect on the
median trend of varying by a factor of two (above and below) all halo
masses around the M+18 relation. Varying halo masses by a factor of
four in this manner has little effect on the model grid and, as a
consequence, on the predicted masses. This is a result of the weak
dependence on halo virial mass of the dark mass enclosed in the inner
regions discussed in Sect.~\ref{SecVcReSims}.

Predicted stellar masses are more sensitive to the halo `contraction' model
assumed to take into account the halo response to the assembly of the
galaxy. Our fiducial model assumes the \citet{Gnedin2004} contraction,
which has been shown to reproduce fairly accurately the results of
cosmological hydrodynamical simulations \citep[see
e.g.][]{Schaller2015b,Cautun2020}. The `adiabatic' contraction
model of \citet{Blumenthal1986}, on the other hand, yields lower
predicted masses and better agreement overall with the observed $M_*$,
as shown by the blue thick line in the right-hand panel of
Fig.~\ref{FigMstrMpredObs}. Indeed, the scatter about the median
`adiabatic' curve is just $0.17$ dex.

The scatter in the predicted versus true stellar mass quoted above is
actually comparable to the mass scatter about the median TFR shown in the left-hand panel of Fig.~\ref{FigScalingObs},
which is $0.28$ dex for the combined P+07 and SPARC samples. This
implies that the grid model in Fig.~\ref{FigMstrMpredObs} may be used,
in principle, as a competitive secondary-distance indicator applicable
to galaxies of all morphologies and that only relies on
distance-independent kinematic and surface brightness
measures. Further work would be needed to express this idea in
explicit observational terms, a task that we defer to future work.

\subsection{The `tilt' of the observed fundamental plane}

\begin{figure}
    \includegraphics[width=\linewidth]{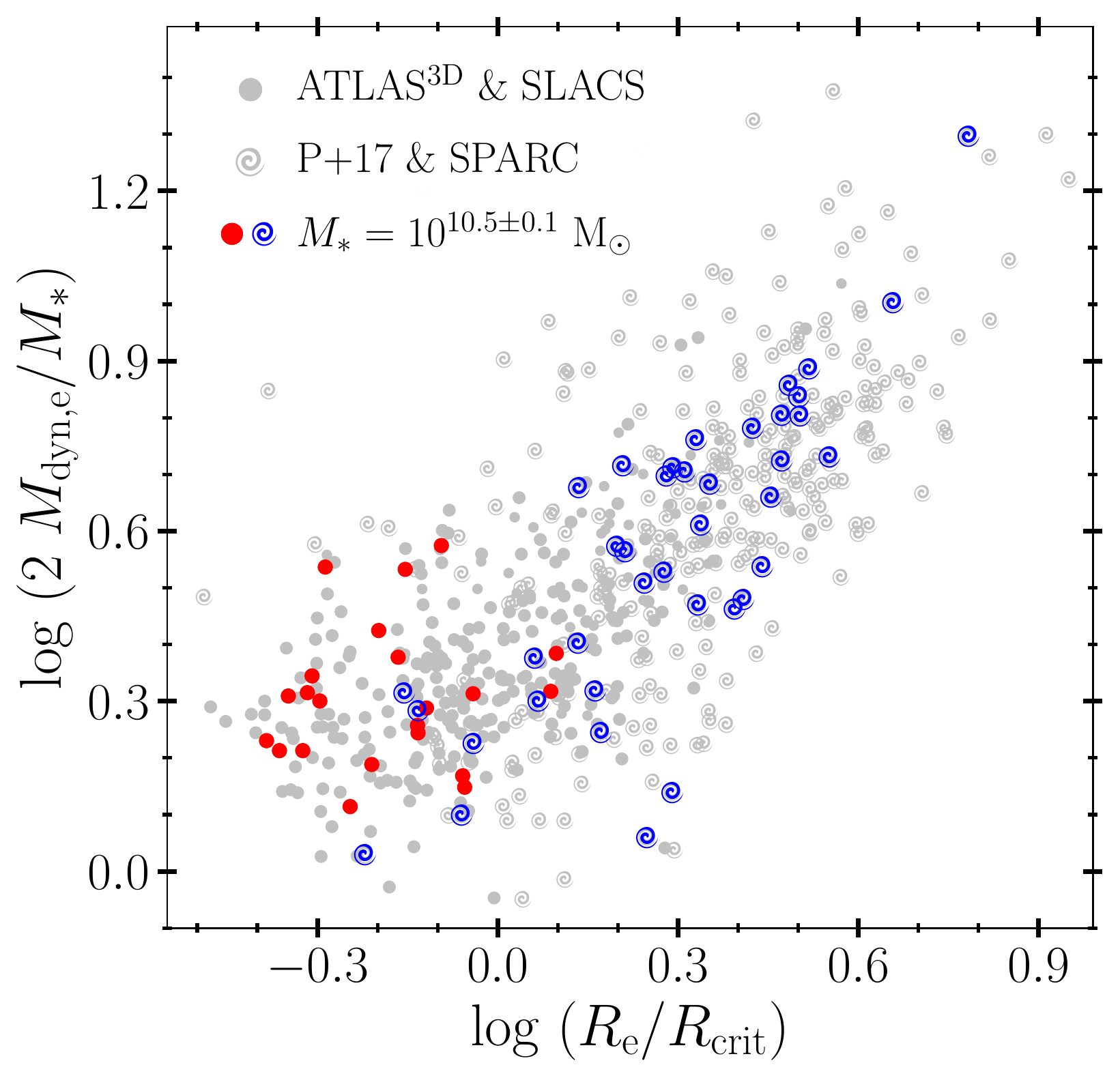}
    \centering
    \caption{Same as Fig.~\ref{FigTiltSims}, but for
        observed galaxies in our sample. Critical radii are computed
        as a function of $M_*$ using our fiducial model and the M+18
        $M_*$-$M_{200}$ relation. Grey symbols denote all observed
        galaxies in our analysis while red and blue correspond to
        S and E galaxies, respectively, with stellar
        masses $M_{*}=10^{10.5 \pm 0.1}$~M$_{\odot}$. It should be noted that the
        dynamical mass-to-light ratio increases with increasing galaxy
        size, at fixed $M_*$. The mass-to-light ratio ratios also
        increase, in general, with increasing galaxy stellar mass
        because massive ellipticals tend to have larger values of
        $R_{\rm e}/R_{\rm crit}$ than less massive early-type systems.
        As for simulated ellipticals, the `tilt' of the FP appears correlate strongly with galaxy size.}
    \label{FigTiltObs}
\end{figure}

We consider next whether the `tilt' in the FP of E galaxies correlates with the size of the galaxy, expressed
in terms of the `critical radius' introduced in
Sect.~\ref{SecTiltSims}. This is shown in
  Fig.~\ref{FigTiltObs}, which indicates that, as for simulated
  galaxies, the `dynamical mass-to-light ratio' of observed galaxies
  does indeed correlate with $R_{\rm e}$/$R_{\rm crit}$, although with
  larger scatter. At fixed $M_*$, galaxies with larger effective radii
  enclose more dark matter than smaller galaxies and tend to have
  higher values of $M_{\rm dyn,e}/(M_*/2)$. As in
  Fig.~\ref{FigTiltSims}, there is a smooth transition between spirals
  and ellipticals at fixed $M_*$. This result offers a simple and
  intuitive explanation for the origin of the `tilt' in the
  FP, as discussed in Sect.~\ref{SecTiltSims}.

\subsection{Comparison between simulated and observed scaling laws}
\label{SecCompSimsObs}

\begin{figure*}
\includegraphics[width=0.75\linewidth]{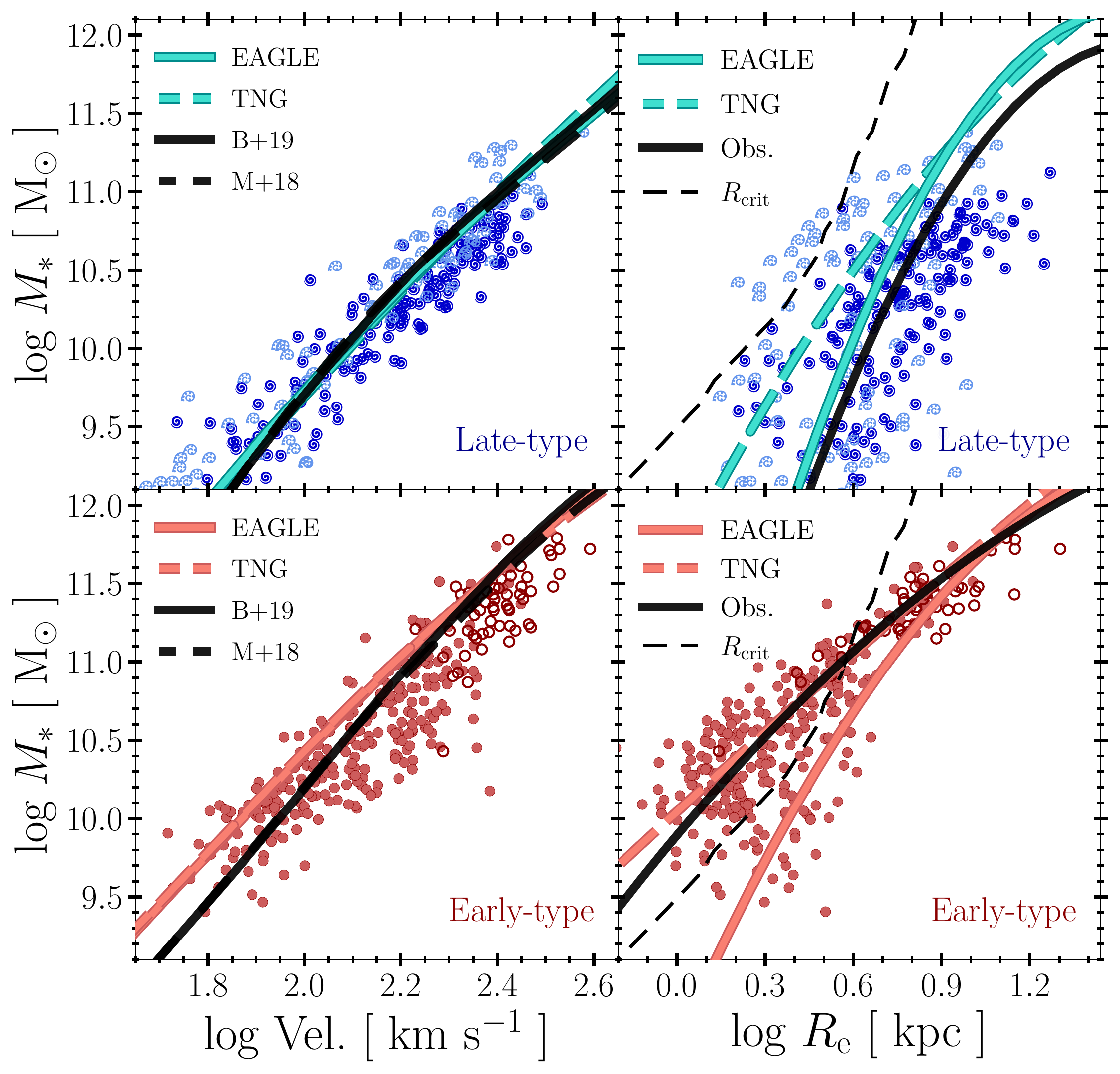}
\centering
\caption{Comparison of observed and simulated galaxy scaling laws. Top
  row corresponds to late-type systems; bottom row to early types. In
  each panel, symbols denote observed galaxies. The smooth coloured
  lines represent the median trend of the simulation results, as in
  Fig.~\ref{FigScalingSims}. it should be noted that simulations match reasonably
  well the rotation velocity-stellar mass (TFR; top left-hand panel).
  The bottom left-hand panel shows line-of-sight velocity dispersions for
  E galaxies versus stellar mass (FJR). Simulated galaxies have
  systematically lower ($\sim 0.1$ dex) velocities than observed
  ellipticals. Thick black lines correspond to the fiducial model
  applied to the hypothetical simulation discussed in
  Sect.~\ref{SecCompSimsObs}. The thin black line in the right-hand
  panels indicate the `critical' radius as a function of $M_*$ for
  the hypothetical simulation. See the discussion in text.}
    \label{FigScalSimsObs}
  \end{figure*}

We end our discussion by comparing the scaling laws of observed and
simulated galaxies. These are shown in Fig.~\ref{FigScalSimsObs},
where the top and bottom rows display, as in Fig.~\ref{FigScalingObs},
the observed correlations between galaxy radius, stellar mass, and
characteristic velocity, split by morphological type (top row for
spirals, bottom row for ellipticals).

The smooth coloured lines show the results of our fiducial model
applied to each of the simulations.  The models use as input, for each
simulation, fits to (i) the galaxy stellar mass-halo mass relation
(Fig.~\ref{FigMgalM200}); (ii) the stellar mass-effective radius
relation (right-hand panels of Fig.~\ref{FigScalingSims}); and (for
ellipticals) (iii) the relation needed to translate
$V_{\rm c,e}$ into $\sigma_{\rm e}$ (Fig.~\ref{FigVcSigma}). Details
on these fits are provided in App.~\ref{SecFits}.

The results of the model are shown by the smooth coloured curves in the left and
middle columns of Fig.~\ref{FigScalingSims}, where they are seen to
provide an excellent description of the simulation results. Because of
this agreement, and to keep the discussion simple, we shall use these models to
discuss the comparison of the simulations with observed data in
Fig.~\ref{FigScalSimsObs}.

The top left-hand panel of Fig.~\ref{FigScalSimsObs} shows that both
simulations do a reasonably good job at reproducing the observed
TFR. Following our earlier discussion, this
agreement is actually expected, and results because the half-mass radii
of simulated spirals are comparable or larger
than their `critical radius'. In this regime,
$V_{\rm c,e}$ becomes insensitive to galaxy radii, and the TFR depends
solely on the $M_*$-$M_{200}$ relation. The good agreement between
observed and simulated TFR thus indicates that TNG and EAGLE spirals
are `large enough', and inhabit halos of the `right' mass \citep{Ferrero2017}.

The comparison of the simulated FJR with observed ellipticals is shown
in the bottom left-hand panel of Fig.~\ref{FigScalSimsObs}. The median
trend of both simulations (coloured lines) is similar, but they appear
to be offset, on average, by $\sim 0.1$ dex in velocity from the
observed FJR. This is intriguing, especially given the excellent
agreement between the simulated and observed TFRs.

The reasons for the FJR offset are easier to appreciate by using the same fiducial model to predict the scaling laws expected for a
hypothetical simulation that reproduces (i) the
abundance-matching galaxy stellar mass-halo mass relation of either M+18 or
B+19 (Fig.~\ref{FigMgalM200}); (ii) the observed galaxy
mass-size relations of ellipticals or spirals (see the solid black curve fits
in the right-hand panels of Fig.~\ref{FigScalSimsObs}); and (iii) the
\citet{Wolf2010} $V_{\rm c,e}$-$\sigma_{\rm e}$ relation (solid line
in Fig.~\ref{FigVcSigma}).

The resulting model Tully-Fisher and Faber-Jackson relations are shown
with black curves in the left-hand panels of
Fig.~\ref{FigScalSimsObs}. It is clear from these panels that the
hypothetical simulation improves the agreement with observation in
both cases. This suggests that the offset in the FJR is related to
differences in at least one of the three model inputs listed above between
the hypothetical simulation and TNG or EAGLE.

For example, compared to the hypothetical simulation model, low-mass
EAGLE ellipticals (say, $M_* \sim 10^{10}\, \rm{M}_\odot$) have similar
halo masses but much larger sizes than observed. Massive EAGLE ellipticals
(say, $M_* \sim 10^{11.5}\, \rm{M}_\odot$), on the other hand, have
similar sizes but much lower halo masses. Either or both of these effects
lead to lower velocity dispersions, which explains the offset between the
observed and simulated FJRs over the whole mass range. Although the
details differ, similar reasoning may also be used to explain
readily the FJR offset of TNG ellipticals.

Interestingly, for massive ellipticals, this analysis leads to a
conclusion that applies to both simulations. In this regime, the FJR
offset arises both in TNG and EAGLE because massive galaxies form in
halos less massive than suggested by abundance-matching
models. Indeed, as may be seen from Fig.~\ref{FigMgalM200}, an
elliptical galaxy with stellar mass
$M_* \sim 3 \times 10^{11} \rm{M}_{\odot}$ forms in a halo with virial
mass $ M_{200} \sim 5\times 10^{13}\, \rm{M}_\odot$ in EAGLE, or
$M_{200} \sim 3\times 10^{13}\, \rm{M}_\odot$ halo in TNG. Massive
simulated galaxies thus form in halos less massive than expected from
AM relations, which predict halos ar least as massive as
$\sim 10^{14}\, \rm{M}_\odot$ for such galaxies. The high velocity
dispersion of massive ellipticals thus suggest that simulations like
EAGLE or TNG need further adjustments in order to prevent massive
galaxies from forming in halos less massive than suggested by
abundance-matching analysis.

Our study is not the first to highlight the differences between observed
and simulated ellipticals. \citet{Lu2020}, for example, have already argued that
fitting in detail the FP of ellipticals requires some
modifications to simulations like TNG. Our analysis confirms and
extends their conclusion, and suggests that only simulations that
match closely and simultaneously the abundance-matching
$M_*$-$M_{200}$ relation as well as the observed sizes of galaxies of
different morphologies will be able to match their observed scaling
laws.\\

\section{Summary and conclusions}
\label{SecConc} 

We have used $\Lambda$CDM cosmological hydrodynamical simulations from the
IllustrisTNG and EAGLE projects to study the scaling laws that relate galaxy
stellar mass with the characteristic size and velocity of luminous
($M_*>10^{10}\, \rm{M}_\odot$) galaxies of different
morphologies. These simulations evolve volumes large enough to yield
statistically significant samples of luminous galaxies and have
numerical resolution high enough to allow for the identification of
galaxies of different morphologies (including rotation-dominated
`spirals' and dispersion-dominated `ellipticals') and for the
measurement of their characteristic sizes and velocities.

The simulations, partly by design, match approximately the galaxy
mass-halo mass relation predicted from abundance-matching analysis
such as those of \citet{Moster2018} and
\citet{Behroozi2019}.
Simulated ellipticals and spirals show little
difference in this regard; at given $M_*$, both morphological types
inhabit halos of similar virial mass. Differences in their scaling
laws thus result mainly from differences in size (ellipticals are
systematically smaller than spirals) and, consequently, in dark matter
content (smaller galaxies enclose less dark mass within their
effective radii).

Our analysis indicates that, for given $M_*$, there is a `critical'
radius that separates galaxies whose characteristic velocities simply
trace the characteristic velocity of their surrounding halo  (i.e. those with
$R_{\rm e}>R_{\rm crit}$, typically spirals) from those where mass and
size play a role in setting a galaxy's characteristic velocity
(i.e. those with $R_{\rm e}<R_{\rm crit}$, typically ellipticals).

This suggests a simple interpretation for why the TFR of spiral galaxies is independent of galaxy radius or surface
brightness, and for the `tilt' of the FP. The latter,
in particular, results from the non-negligible dark matter
contribution to dynamical mass estimates. The dark matter content
depends directly on the ratio $R_{\rm e}/R_{\rm crit}$, implying that
the FP tilt is primarily driven by galaxy `size' rather than mass.

A simple fiducial model that combines the NFW-like structure of $\Lambda$CDM
halos with the $M_*$-$M_{200}$ relation of each simulation is able to
reproduce well the scaling laws of simulated galaxies. The same
fiducial model suggests how scaling laws may be unified into a simple
scenario where galaxy stellar mass is determined uniquely by its
effective radius, $R_{\rm e}$, and its circular velocity at the
stellar half-mass radius, $V_{\rm c,e}$, regardless of morphology.

The model depends only on the assumed $M_*$-$M_{200}$ relation, and
implies that a secondary distance indicator may be constructed by
combining two distance independent quantities, the circular velocity,
$V_{\rm c,e}$, and the effective surface brightness, $\Sigma_{\rm e}$,
to predict the total stellar mass of a galaxy. This distance indicator
is remarkably precise, with little bias and a stellar mass
(luminosity) scatter of only $0.076$ dex for TNG and $0.078$ dex for
EAGLE.

The analysis discussed above for TNG and EAGLE may also be applied to
observational data to provide a cosmological interpretation for the
observed galaxy scaling laws and to elucidate the
reasons for differences that may arise when confronting simulations
with observation. Overall, observed scaling laws seem broadly consistent with the
interpretive framework proposed by the simulations. In particular, the
interpretation of both the `tilt' of the FP and of
the surface brightness independence of the TFR as
related to galaxy size seem consistent with the data. Even the unified
$V_{\rm c,e}$-$\Sigma_{\rm e}$ distance indicator, applied to a galaxy
sample that includes galaxies of all morphological types, gives
competitive results, with little bias and a stellar mass (luminosity)
scatter of $0.17$-$0.19$ dex.

Comparing observations with simulations, the few differences that
arise can be traced to deviations from either the observed galaxy
mass-size relations, or from the abundance-matching galaxy stellar mass-halo
mass relation.  A hypothetical $\Lambda$CDM simulation where those two
requirements are closely and simultaneously met seems broadly
consistent with  the relations linking the size,
mass, and kinematics of galaxies of all morphological types. This
should be rightly regarded as a major success of the $\Lambda$CDM model in the
highly non-linear regime of the inner regions of individual galaxies.

\section*{Acknowledgements}
\label{sec:ackno} 

We thank Laura Sales for useful comments and Isabel Santos Santos for
providing the SPARC data used in this study in electronic form.  MGA,
JAB and IF acknowledge financial support from CONICET through PIP
11220170100527CO grant.



\bibliographystyle{aa.bst}
\bibliography{references} 

\appendix

\section{Morphological classification of simulated galaxies}
\label{SecMorphClass}

\begin{figure*}
    \includegraphics[width=0.75\linewidth]{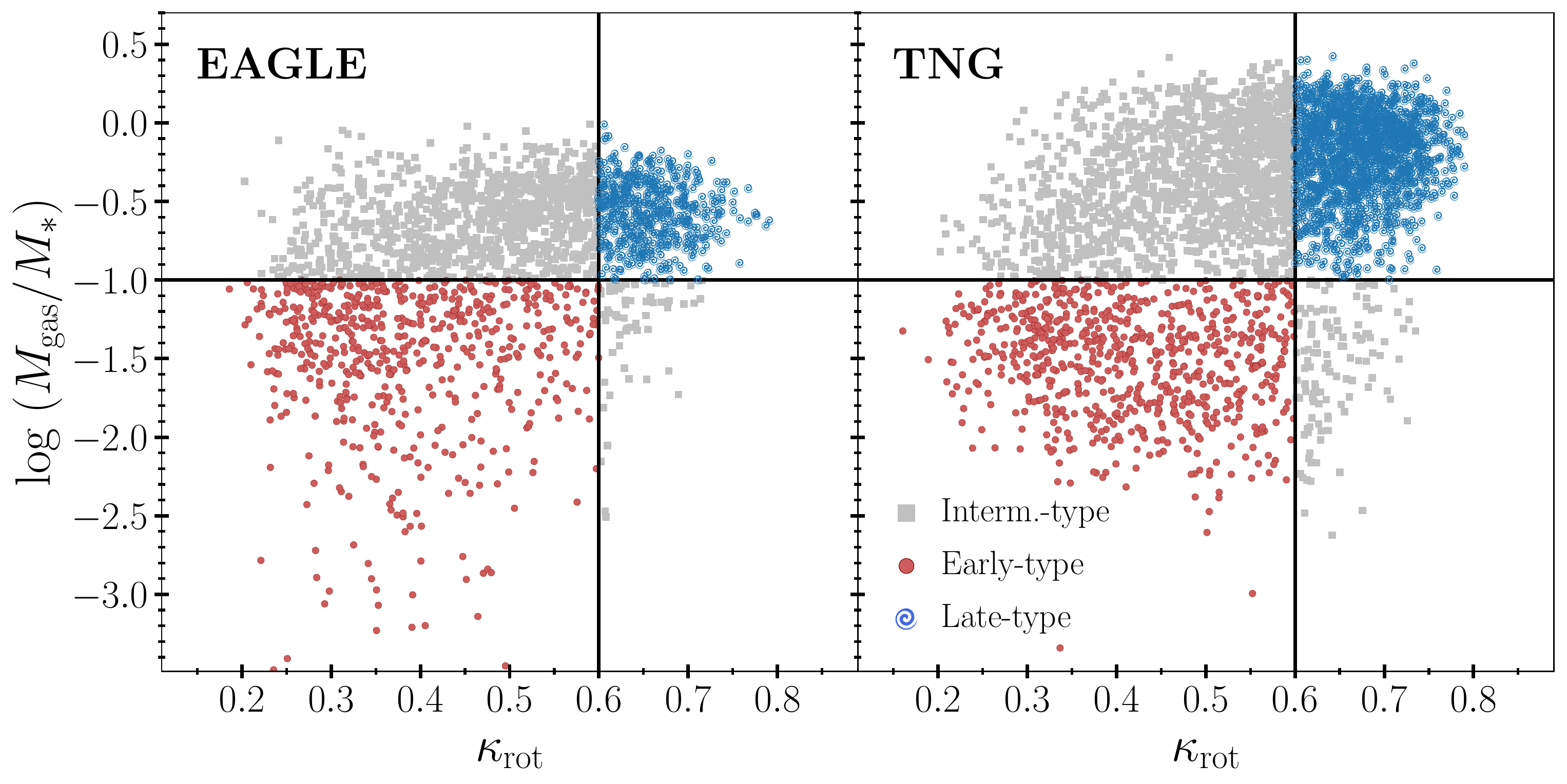}
\centering
\caption{Gas mass fraction within $r_{\rm gal}$ as a function of the 
rotational-to-total kinematic energy parameter, $\kappa_{\rm rot}$. The left-hand panel corresponds to 
EAGLE, right-hand panel to TNG. Symbols show individual galaxies coloured by 
morphology: Red circles indicate early-type (E); blue spiral symbols denote
 late-type (S); grey squares are used for all others. Vertical and horizontal lines denotes
 cutoff values for morphological classification: $\kappa_{\rm rot}=0.6$ and $M_{\rm gas}/M_{*} = 0.1$, respectively. }
    \label{FigMorphClass}
\end{figure*}

Motivated in part by the fact that our main conclusions are applicable
to all simulated galaxies, regardless of morphology, we adopt here a
very simple classification procedure that takes into account only the
rotational support of the stellar component and the gas content of a
galaxy. The first parameter gauges the presence of a stellar disc and
its relative prominence; the second parameter is a proxy for recent or
ongoing star formation, which also typically occurs in a disc.

Rotational support is measured by the fraction of kinetic energy
invested in ordered rotation \citep{Sales2012}:
\begin{equation}
\label{krot}
\kappa_{\rm rot} = \frac{K_{\rm rot}}{K} = \frac{1}{K} \sum \frac{1}{2}~m\left ( \frac{j_{z}}{R} \right )^{2} .
\end{equation}
The parameter $\kappa_{\rm rot}$ approaches unity for systems with
perfect circular motions (disc-dominated) and approaches zero for
non-rotating pressure supported spheroidal systems. This parameter is
a simple quantitative measure of morphology and correlate extremely
well with the fraction of stars with circularity parameters
$\epsilon_{j}>0.5 $ (commonly used criterion for morphological
classification). This parameter was exhaustively analysed in
\citet{Ferrero2012}, where a cutoff value of $\kappa_{\rm rot} = 0.6$
was found to best separate discs from spheroids.

The second parameter is the ratio between gas and stellar mass within
the galactic radius, $r_{\rm gal}$, $f_{\rm gas}=M_{\rm gas}/M_{*} $.
The cutoff value adopted for this parameter is $f_{\rm gas}=
0.1$. This parameter correlates well with the specific star formation
rate, sSFR, in both simulations.On average a galaxy with
$f_{\rm gas}=0.1$ has $log~$sSFR$~=-10.5~$yr$^{-1}$. We decided to use
the gas mass fraction instead of the sSFR because using the latter
yields a large number of E galaxies with
low sSFR but substantial amounts of gas.

To summarize, the morphological classification is performed by combining
the kinematic parameter $\kappa_{\rm rot}$
with the content parameter $M_{\rm gas}/M_{*}$. Galaxies
with $\kappa_{\rm rot} < 0.6$ and $M_{\rm gas}/M_{*}  < 0.1$ are flagged as early-type. On the other hand, late-type galaxies are those with $\kappa_{\rm rot} > 0.6$ and $M_{\rm gas}/M_{*}  > 0.1$.

Figure~\ref{FigMorphClass} shows the gas mass fraction as a function 
of  $\kappa_{\rm rot}$, EAGLE in the left-hand panel and TNG in
the right. Blue spiral symbols and red circles denote galaxies 
selected as late-type and early-type, respectively. Grey squares 
indicate galaxies that do not meet both parameters at the same time, 
which we label as intermediate
types.\\

\section{Model fits}
\label{SecFits}

The fits to the median trends of the stellar mass $M_*$, as a function
of the half-mass radius, $R_e$ shown in the right-hand panels of
Fig.~\ref{FigScalSimsObs} follow a polynomial of degree two:
log$~M_* = a~$log$~R_{\rm e}^2 + b~$log$~R_{\rm e} + c$. The values of $a$, $b$ and
$c$ for, simulated and observed S and E galaxies may be
found in Table~\ref{tab:my-table}.

For the galaxy mass-halo mass relation, $M_*(M_{200})$, for
observations we asume fits to the M+18 or B+19 (solid and dashed black
lines on Fig.~\ref{FigMgalM200}) curves following the function
proposed in \citet{Moster2010}:
\begin{equation}
	\frac{M_*}{M_{200}} = 2 A \left[ \left(\frac{M_{200}}{M_1}\right)^{-\beta} 
	+ \left(\frac{M_{200}}{M_1}\right)^{\gamma} \right]. 
\end{equation}
Table~\ref{tab:my-table} shows the values of $A$, $M_1$, $\beta$ and
$\gamma$ appropriate for the EAGLE and TNG simulations. For clarity, these fits are not shown in
Fig.~\ref{FigMgalM200}, but they follow the corresponding open symbols
in that figure.

%
{\renewcommand{\arraystretch}{1.6}%
\begin{table}
\large
\centering
\caption{Parameter values for fits used for $M_* (M_{200})$ and log$~M_* ($log$~R_{\rm e})$.
All masses are in units of M$_{\odot}$.}
\resizebox{\linewidth}{!}{%
\begin{tabular}{|c|c|l|c|c|c|l|c|c|c|c|}
\cline{1-2} \cline{4-6} \cline{8-11}
\multirow{2}{*}{Sample} &
  \multirow{2}{*}{Type} &
   &
  \multicolumn{3}{c|}{log$~M_* ($log$~R_{\rm e})$} &
   &
  \multicolumn{4}{c|}{$M_* (M_{200})$} \\ \cline{4-6} \cline{8-11} 
 &   &  & $a$     & $b$    & $c$     &  & $A$ & $M_1$ & $\beta$ & $\gamma$ \\ \cline{1-2} \cline{4-6} \cline{8-11} 
\multirow{2}{*}{EAGLE} &
  S &
   &
  $-2.43$ &
  $7.44$ &
  $6.47$ &
   &
  \multirow{2}{*}{$0.0193$} &
  \multirow{2}{*}{$1.63 \times 10^{12}$} &
  \multirow{2}{*}{$0.464$} &
  \multirow{2}{*}{$0.602$} \\ \cline{2-2} \cline{4-6}
 & E &  & $-1.18$ & $4.24$ & $8.56$  &  &     &       &         &          \\ \cline{1-2} \cline{4-6} \cline{8-11} 
\multirow{2}{*}{TNG} &
  S &
   &
  $-0.79$ &
  $3.61$ &
  $8.62$ &
   &
  \multirow{2}{*}{$0.29$} &
  \multirow{2}{*}{$1.12 \times 10^{12}$} &
  \multirow{2}{*}{$0.477$} &
  \multirow{2}{*}{$0.499$} \\ \cline{2-2} \cline{4-6}
 & E &  & $-0.06$ & $1.72$ & $10.04$ &  &     &       &         &          \\ \cline{1-2} \cline{4-6} \cline{8-11} 
\multirow{2}{*}{Obs.} &
  S &
   &
  $-2.27$ &
  $7.13$ &
  $6.34$ &
   &
  \multicolumn{4}{c|}{\multirow{2}{*}{M+18 / B+19}} \\ \cline{2-2} \cline{4-6}
 & E &  & $-0.46$ & $2.23$ & $9.9$   &  & \multicolumn{4}{c|}{}            \\ \cline{1-2} \cline{4-6} \cline{8-11} 
\end{tabular}%
}
\label{tab:my-table}
\end{table}
}

Finally, we provide an analytical expression for the
critical radius, $R_{\rm crit}$ dependence on $M_*$. To model $R_{\rm crit}$ a relation between 
stellar mass and halo mass as well as a halo contraction model need to be assumed. Table~\ref{table2} shows the parameters of a 
polynomial fit of degree two: log$~R_{\rm crit}= a~$log$~M_{*}^2 +
b~$log$~M_{*} + c$. The fits shown in Table~\ref{table2} assume a \citet{Gnedin2004} contraction and are valid for $M_{*}>10^{9.5}$~$M_{\odot}$.\\  

As an extra resource for the interested reader we provide on
\href{https://github.com/ferreroismael/DistanceIndicator.git}
{GitHub\footnote{https://github.com/ferreroismael/DistanceIndicator.git}} a code that
builds the grids shown in Fig.~\ref{FigMstrMpredSims} and Fig.~\ref{FigMstrMpredObs}.
More generally, the code returns the predicted stellar mass by providing the surface
brightness, $\Sigma_{\rm e}$, and the effective radii, $R_{\rm
  e}$. The code also allows  different stellar to halo mass relations (EAGLE, TNG, M+18 or B+19)
and different models for the halo contraction \citep[][ or without contraction]{Gnedin2004,Blumenthal1986}.

{\renewcommand{\arraystretch}{1.6}%
\begin{table}
\large
\centering
\caption{Parameter values for fits used for log$~R_{crit}( $log$~M_{*})$}
\resizebox{0.6\linewidth}{!}{%
\begin{tabular}{|c|l|l|c|c|c|}
\cline{1-2} \cline{4-6}
\multicolumn{2}{|c|}{\multirow{2}{*}{Sample}} &  & \multicolumn{3}{c|}{log$~R_{crit}( $log$~M_{*})$} \\ \cline{4-6} 
\multicolumn{2}{|c|}{}      &  & a      & b     & c      \\ \cline{1-2} \cline{4-6} 
\multicolumn{2}{|c|}{EAGLE} &  & -0.069 & 1.819 & -11.05 \\ \cline{1-2} \cline{4-6} 
\multicolumn{2}{|c|}{TNG}   &  & -0.065 & 1.765 & -10.88 \\ \cline{1-2} \cline{4-6} 
\multicolumn{2}{|c|}{M+18}  &  & -0.072 & 1.856 & -11.13 \\ \cline{1-2} \cline{4-6} 
\multicolumn{2}{|c|}{B+19}  &  & -0.071 & 1.845 & -11.23 \\ \cline{1-2} \cline{4-6} 
\end{tabular}%
}
\label{table2}
\end{table}
}

\label{lastpage}
\end{document}